\documentclass{article}
\usepackage[letterpaper,textwidth=5.5in,textheight=9in,top=1in,headheight=12pt,headsep=25pt,footskip=30pt]{geometry}
\usepackage{times}
\usepackage{natbib}
\setcitestyle{authoryear,round,citesep={;},aysep={,},yysep={;}}
\usepackage[utf8]{inputenc}
\usepackage[T1]{fontenc}
\usepackage{hyperref}
\usepackage{url}
\usepackage{booktabs}
\usepackage{amsfonts}
\usepackage{amssymb}
\usepackage{nicefrac}
\usepackage{microtype}
\usepackage[table]{xcolor}
\usepackage{array}

\colorlet{heatnan}{gray!20}

\usepackage{graphicx}
\usepackage{latexml}
\iflatexml
  \renewcommand{\resizebox}[3]{#3}
  \renewcommand{\rotatebox}[3][]{#3}
\else
  \expandafter\def\expandafter\UrlBreaks\expandafter{\UrlBreaks
    \do\a\do\b\do\c\do\d\do\e\do\f\do\g\do\h\do\i\do\j\do\k\do\l\do\m
    \do\n\do\o\do\p\do\q\do\r\do\s\do\t\do\u\do\v\do\w\do\x\do\y\do\z}
\fi
\usepackage{enumitem}
\usepackage{pifont}
\usepackage{multirow}
\usepackage{colortbl}

\usepackage{xspace}
\usepackage{tabularx}
\usepackage{ragged2e}
\usepackage{makecell}

\newcolumntype{R}[1]{>{\raggedleft\arraybackslash}p{#1}}
\usepackage{amsmath}

\newcommand{\voicenet}{\textsc{VoiceNet}\xspace}
\newcommand{\emonet}{\textsc{EmoNet-Voice}\xspace}
\newcommand{\emonetface}{\textsc{EmoNet-Face}\xspace}
\newcommand{\voicenetemo}{\textsc{VoiceNet-Emo}\xspace}
\newcommand{\voicenetext}{\textsc{VoiceNet-Ext}\xspace}
\newcommand{\emolia}{\textsc{Emolia}\xspace}
\newcommand{\emoliabal}{\textsc{Emolia-Balanced}\xspace}
\newcommand{\voiceclap}{\textsc{VoiceCLAP}\xspace}
\newcommand{\voiceclapsmall}{\textsc{VoiceCLAP-Small}\xspace}
\newcommand{\voiceclaplarge}{\textsc{VoiceCLAP-Large}\xspace}
\newcommand{\voiceclapdata}{\textsc{VoiceCLAP Data}\xspace}

\title{\voicenet: Fine-Grained Voice Understanding Beyond Emotion at Scale}

\author{%
  Christoph Schuhmann$^{*1}$ \quad Robert Kaczmarczyk$^{*1,2}$ \quad Gollam Rabby$^{3}$ \\
  Felix Friedrich$^{4}$ \quad Maurice Kraus$^{5}$ \quad Gijs Wijngaard$^{2}$ \quad Kourosh Nadi$^{1}$ \\
  Huu Nguyen$^{6}$ \quad Kristian Kersting$^{5,7,8}$ \quad S\"oren Auer$^{3,9}$ \\[8pt]
  \small $^{1}$LAION e.V. \enspace
  $^{2}$Scalable Learning \& Multi-Purpose AI (SLAMPAI) Lab, Forschungszentrum J\"ulich GmbH \\
  \small $^{3}$L3S Research Center \enspace
  $^{4}$Black Forest Labs \enspace
  $^{5}$Computer Science Department, TU Darmstadt \\
  \small $^{6}$Ontocord.AI \enspace
  $^{7}$German Research Center for Artificial Intelligence (DFKI) \\
  \small $^{8}$Hessian Center for AI (hessian.AI) \enspace
  $^{9}$Leibniz Universit\"at Hannover \\[6pt]
  \texttt{contact@laion.ai} \\[4pt]
  \small $^{*}$Equal contribution (shared first authors)
}
\date{}

\begin{document}

\maketitle

\begin{abstract}
  Expressive speech synthesis has outpaced expressive speech perception: systems now render fine-grained vocal performances that no public benchmark can score. Most benchmarks for this inverse problem stop at six to nine basic emotion categories, largely on acted speech. This paper introduces \voicenet, a human-annotated representation-level benchmark for voice performance understanding on permissively-licensed in-the-wild speech. \voicenet has two subsets: \voicenetemo applies a 40-emotion taxonomy with three expert ratings per item, and \voicenetext, a preliminary subset, scores 57 talking-style attributes including speaking rate, vocal tension, breathiness, and register. The paper also releases \emolia, an emotion-annotated version of the Emilia corpus, with a curated rebalanced subset enriched by dense MOSS-Audio Thinking annotations. Two voice-text contrastive baselines train on this data: a 110M-parameter \voiceclapsmall for fast large-scale data filtering and a 7B \voiceclaplarge for state-of-the-art performance. Both outperform existing CLAP baselines, which sit near chance on \voicenetemo. On \voicenetemo, \voiceclaplarge aligns more closely with the aggregate expert consensus than individual experts agree with one another: a comparison against the majority label rather than evidence of surpassing human emotion perception. All systems evaluated here are voice-text embedding models: \voicenet scores representation-level attribute recognition and retrieval, not end-to-end spoken-dialogue behaviour. Clustering and filtering uncurated speech corpora into subsets that span diverse talking styles and emotions remains an open challenge; \voiceclap embeddings offer a promising tool for this task. \voicenet, \emolia, and \voiceclap are publicly available for research use.
\end{abstract}

\section{Introduction}
\label{sec:introduction}

Synthetic speech technology generates fluent, expressive speech \citep{hurst2024gpt,Kirk2025why}, but recognizing fine-grained vocal performance has not kept pace \citep{cowie2001emotion,schuller2018voice}; two evaluation gaps slow progress. First, public speech-emotion benchmarks rely on a small set of basic emotions \citep{ekman1992argument,george2024review}: \textsc{IEMOCAP} \citep{busso2008iemocap}, \textsc{RAVDESS} \citep{livingstone2018ravdess}, \textsc{CREMA-D} \citep{cao2014crema}, \textsc{SAVEE} \citep{jackson2014savee}, and \textsc{EmoDB} \citep{burkhardt2005emodb} cover 6 to 9 emotions on studio-acted speech, and aggregation efforts \citep{scheidwasser_clow_et_al_2021,ma_et_al_2024,osman_et_al_2024} inherit those limits. Second, no public benchmark scores talking-style attributes (speaking rate, vocal tension, breathiness, register, chunking, timbre, recording context) alongside emotion, despite their established role in paralinguistic perception \citep{schuller2013-kn,cowie2001emotion}. Synthetic resources such as \emonet \citep{schuhmann2026emonetvoicefinegrainedexpertverifiedbenchmark} address taxonomy granularity but use text-to-speech audio; whether models trained on dense annotations transfer to in-the-wild speech is open.

Three observations motivate this paper. The 40-emotion taxonomy grounded in modern affective science \citep{barrett2017theory,russell1980circumplex,cowen2019mapping} has been validated on synthetic \emonet audio and applies to permissively-licensed real-human speech. Existing CLAP \citep{wu2023clap,elizalde2024msclap} and voice-text models \citep{dinkel2026glap,yang2026towards} are trained on general-audio captions, not dense vocal-style descriptions. Open-source audio language models such as \textsc{MOSS-Audio-8B-Thinking} \citep{mossaudio2026} produce structured per-clip annotations across emotion, voice quality, and acoustic context at scale.

This paper contributes:
(1) \textbf{\voicenet benchmark.} A human-annotated benchmark on permissively-licensed in-the-wild speech with two subsets: \voicenetemo extends the \emonet 40-emotion taxonomy to 7,988 (audio, prompt) pairs over 3,944 clips with three expert raters per item; \voicenetext covers 57 talking-style attributes\footnote{\url{https://projects.laion.ai/emolia-bench/taxonomy/}} over 18.5k (audio, level) pairs carrying 40{,}990 ratings from eight annotators.
(2) \textbf{\emolia, \emoliabal, and the MOSS-Audio annotation suite.} A fully emotion-annotated version of Emilia (71.78M clips with 40 emotion scores, captions, and speaker embeddings), a curated 5.26M-clip subset (\emoliabal) rebalanced across 40 emotions and 3,000 speaker clusters, and \textsc{MOSS-Audio-8B-Thinking} annotations on \emoliabal plus three open voice corpora (8.64M clips, 18 prompt-groups, 61 attribute values per clip).
(3) \textbf{\voiceclap models.} \voiceclapsmall is a 110M-parameter dual-tower BUD-E-Whisper-Small + all-MiniLM-L6-v2 trained with SigLIP. \voiceclaplarge is a rank-16 LoRA finetune of \textsc{LCO-Embedding-Omni-7B} \citep{xiao2025scaling} trained with InfoNCE. \voiceclaplarge reaches 0.702 per-prompt balanced accuracy on \voicenetemo (+0.039 over the zero-shot Omni base).
(4) \textbf{Evaluation.} Seven general-audio CLAPs \citep{wu2023clap,elizalde2024msclap,dinkel2026glap,yang2026towards,niizumi2025m2d2,li2024advancing,zhu2024cacophony} sit at chance on \voicenetemo ($|\rho|<0.1$); two zero-shot Omni-Embedding bases \citep{xiao2025scaling} reach 0.65 to 0.66.

\textbf{Scope.} \voicenet is a \emph{representation-level} benchmark: it measures how well voice-text embedding models rank fine-grained emotion and talking-style descriptions against real speech. Every system evaluated in \S\ref{sec:experiments} is an audio-text embedding model; end-to-end voice assistants, conversational context, response generation, and generative audio-language models (which require a different scoring interface than cosine similarity) are out of scope.

Beyond the headline benchmark, clustering and filtering raw uncurated speech corpora into subsets that span diverse talking styles and emotions remains an open challenge. \voiceclap embeddings provide a practical handle on this problem: nearest-neighbour search on the embedding space surfaces clips by emotion, talking-style, and recording context. An interactive browser of \emoliabal\footnote{\url{https://projects.laion.ai/emolia-bench/demo/}} demonstrates this use case.

\section{Related Work}
\label{sec:related_work}

\textbf{Speech emotion benchmarks.} Public SER datasets cluster around six to nine acted emotion classes: \textsc{IEMOCAP} (12h, 9 emotions) \citep{busso2008iemocap}, \textsc{RAVDESS} (1h, 8) \citep{livingstone2018ravdess}, \textsc{SAVEE} (0.8h, 7) \citep{jackson2014savee}, \textsc{EmoDB} (German, 7) \citep{burkhardt2005emodb}, and \textsc{CREMA-D} (6) \citep{cao2014crema}. Their taxonomies derive from basic-emotion theory \citep{ekman1992argument,plutchik2001nature}, whereas vocal-expression studies map a far larger space of distinct emotions \citep{cowen2019mapping}; acted prosody overstates cues \citep{wilting2006real}, and ethical barriers prevent collection of stigmatising emotions \citep{schuller2013-kn}. Aggregations such as \textsc{SERAB} (9 corpora, 6 languages) \citep{scheidwasser_clow_et_al_2021}, \textsc{EmoBox} (32 datasets, 14 languages) \citep{ma_et_al_2024}, and \textsc{SER Evals} (18 minority-language corpora) \citep{osman_et_al_2024} broaden coverage but inherit acted speech and narrow taxonomies; \textsc{BERSt} adds shouted speech from 98 actors \citep{tuttosi_et_al_2025}, and \textsc{MSP-Podcast} \citep{lotfian2019msp} is one of few in-the-wild corpora at scale. The \emonet bench \citep{schuhmann2026emonetvoicefinegrainedexpertverifiedbenchmark} introduces the 40-emotion taxonomy on synthetic audio; \voicenet ports it to in-the-wild human speech, adds 57 talking-style attributes, and uses contrastive cosine-similarity scoring rather than ordinal regression.

\textbf{Voice-text contrastive models.} Contrastive Language-Audio Pretraining was first applied to general-audio sound events \citep{wu2023clap,elizalde2024msclap}. Subsequent models target singing voice \citep{yang2026towards} or unify speech, music, and sound \citep{dinkel2026glap}. Training data for these models comes from AudioCaps and Clotho, captioned at sound-event level. None target dense vocal-style annotation. \S\ref{sec:experiments} confirms that all general-audio CLAPs sit at chance on \voicenetemo. Single-tower audio-language embedding models (\textsc{LCO-Embedding-Omni-7B} \citep{xiao2025scaling}, \textsc{LCO-Embedding-Omni-3B} \citep{xiao2025scaling}) inherit speech understanding from instruction-tuned multimodal LLMs, reaching approx.\ 0.65 on \voicenetemo zero-shot.

\textbf{Annotation taxonomies and dimensional models.} Affective science models emotions as context-dependent and graded \citep{barrett2017theory,lindquist2013constructionist}. Valence-arousal-dominance \citep{russell1980circumplex} and multi-label schemes \citep{zhang2020multi,cowen2019mapping} support blended affect. Most benchmarks still assign a single discrete label per clip. Intensity annotations show low crowd-source agreement \citep{kajiwara2021wrime,stappen2021muse}. \voicenetext takes a complementary route by annotating 57 talking-style attributes with binary level rubrics, broadening the supervision signal beyond emotion categories.

\section{\voicenet Suite}
\label{sec:dataset}

The suite has four released components: the 40-emotion + 57-attribute taxonomy (\S\ref{sec:taxonomy}); \emolia (the fully annotated Emilia corpus) with \emoliabal and three sibling MOSS-Audio-annotated training corpora (\S\ref{sec:methods_emolia}); the \voicenet human-annotated benchmark with two subsets at different maturity levels the near-complete, expert-rated \voicenetemo and the preliminary \voicenetext, whose inter-rater agreement is at chance (\S\ref{sec:methods_bench}); and the two \voiceclap contrastive models (\S\ref{sec:methods_voiceclap}).

\subsection{Taxonomies}
\label{sec:taxonomy}

\textbf{Emotion taxonomy (40 categories).} The 40-category emotion taxonomy originally developed for \emonetface \citep{schuhmann2025emonetface} covers positive emotions (\textit{Elation}, \textit{Contentment}, \textit{Affection}, and \textit{Awe}), negative emotions (\textit{Distress}, \textit{Sadness}, \textit{Bitterness}, and \textit{Contempt}), cognitive states (\textit{Concentration}, \textit{Confusion}, and \textit{Doubt}), physical states (\textit{Pain}, \textit{Fatigue}), and socially mediated emotions (\textit{Embarrassment}, \textit{Shame}, \textit{Pride}, and \textit{Teasing}). The full set of 40 categories with descriptive terms appears in App.~\ref{app:taxonomy}. The literature-extraction and expert-guided refinement process is detailed in App.~\ref{app:taxonomy_construction}.

\textbf{Talking-style attribute taxonomy (57 attributes).} \voicenetext covers a parallel 57-attribute taxonomy of talking-style and paralinguistic descriptors, grouped into perceived speaker traits, affective dimensions, prosodic delivery, vocal quality, resonance placement, recording context, and style descriptors (full list with short codes in App.~\ref{app:dim_attributes}). The 57 attributes derive from the 61-value \textsc{MOSS-Audio} schema (\S\ref{sec:methods_emolia}) by dropping vocal-burst presence (\textsc{BURST}, redundant with discrete event detection), fine-grained categorical emotion (\textsc{EMO}, covered by \voicenetemo), accent (\textsc{ACNT}, demographic rather than perceptual), and language (\textsc{LANG}, categorical rather than ordinal). The rubric-level definitions used for human annotation are available at \url{https://projects.laion.ai/emolia-bench/taxonomy/}.

\subsection{\emolia, \emoliabal, and the \textsc{MOSS-Audio} Annotation Pipeline}
\label{sec:methods_emolia}

\textbf{\emolia} is a fully emotion-annotated version of the open-source Emilia corpus. The complete Emilia-Large release (71.78M clips, 215{,}600 hours; its Emilia portion is licensed CC-BY-NC-4.0 and its Emilia-YODAS portion CC-BY-4.0) is annotated with 40 emotion scores from the EmpathicInsight-Voice classifier, emotion captions from BUD-E-Whisper, and WavLM-based speaker-timbre embeddings. \emolia is released in full as a large-scale training resource for emotion-aware speech research.

\textbf{\emoliabal} is a curated 5.26M-clip subset of \emolia, rebalanced across the 40 emotions and 3,000 k-means speaker-embedding clusters using empathic-voice-classifier logits. Together with 3 additional open voice corpora (LAION's Got Talent 1.69M clips, Majestrino 973k clips, Multilingual In The Wild 721k clips), \emoliabal forms the MOSS-Audio annotation suite (8.64M clips total; per-corpus counts in App.~\ref{app:emolia_corpora}, Tab.~\ref{tab:emolia_corpora}). An internal ablation excluded Multilingual In The Wild from the training mixture (7.92M clips) without dropping it from the release.

\textbf{Annotation pipeline.} Each clip is queried with 18 prompt-groups covering related attribute clusters (resonance placement, prosodic delivery, recording context, style category, etc.). \textsc{MOSS-Audio-8B-Thinking} \citep{mossaudio2026} returns a structured response per group; the parser consumes only the output after the closing \verb|</think>| tag and regenerates malformed groups. The 18 outputs supply 61 short-code attribute values per clip, concatenated into one \texttt{text} field for contrastive training. The chain-of-thought traces emitted before \verb|</think>| are retained and released alongside the structured annotations (155.46M traces total) to support future work on reasoning-supervised voice understanding; they are not used for the contrastive training in this paper.

\textbf{\emoliabal construction.} The Emilia source corpus is filtered with \emph{EmpathicVoice} released with \emonet bench \citep{schuhmann2026emonetvoicefinegrainedexpertverifiedbenchmark}, a Whisper-based audio encoder with a 40-dim.\ linear head trained on synthetic \emonet-Voice to predict per-clip presence logits. The classifier is used for stratified sampling and benchmark prompt generation, never as ground truth. Clips are sampled to balance the 40-emotion categories and 3,000 k-means clusters of WavLM speaker embeddings. Rebalancing softens the source-corpus emotion skew of Emilia but does not fully remove it.

\textbf{Training corpora.} \voiceclap trains on nine voice corpora: \emoliabal (5.26M), LAION's Got Talent (1.69M), Majestrino (973k), two in-house Synthetic Vocal Bursts collections (341k), and four FCaps-captioned corpora \citep{yang2026towards}: EARS (17k), Expresso (27k), Voxceleb1 (154k), Voxceleb2 (600k). The smaller six broaden the contrastive negative distribution; their captions are not part of the \emolia suite. Full counts in App.~\ref{app:detailed_baseline_training}, Tab.~\ref{tab:voiceclap_data_manifest}.

\subsection{\voicenet: A Human-Annotated In-the-Wild Benchmark}
\label{sec:methods_bench}

\voicenet evaluates fine-grained voice understanding on permissively-licensed in-the-wild human speech. Unlike the prior \emonet bench \citep{schuhmann2026emonetvoicefinegrainedexpertverifiedbenchmark}, which evaluates on synthetic TTS-generated audio, \voicenet consists exclusively of real human vocal performances; the \voicenetemo clips come from the Emilia-YODAS (CC-BY-4.0) and Emilia (CC-BY-NC-4.0) portions of Emilia-Large, so the benchmark is openly available under CC-BY and CC-BY-NC research licences. Each clip is paired with a textual prompt. Raters mark whether the property described by the prompt is present in the clip. Two subsets share this protocol but differ in prompt vocabulary, audio source, and rater pool.

\textbf{\voicenetemo (40 emotions, expert raters).}
\voicenetemo applies the same 40-emotion taxonomy as \emonet but to naturalistic, in-the-wild human speech rather than synthetic audio. It asks: \emph{is emotion $E$ present in clip $C$?} for each of the 40 emotions. The subset contains 7,988 (audio, emotion) pairs over 3,944 unique audio clips and 40 emotion prompts. 7,984 of those pairs received the full three independent ratings from psychology experts on the three-point scale used in prior work \citep{schuhmann2026emonetvoicefinegrainedexpertverifiedbenchmark}: 0 (not present), 1 (weakly present), and 2 (strongly present); the remaining 4 pairs received one or two ratings due to incomplete rater coverage and are excluded from inter-rater statistics but retained in the released benchmark. Three experts contributed (\texttt{user\_0}, \texttt{user\_1}, and \texttt{user\_2}). Each (audio, emotion) pair was rated independently with assignments balanced for rater gender. Annotators were blinded to one another's ratings.

Each pair is sampled under one of five task types so that emotion presence is balanced across the subset. \emph{Affirmative} (4,000 pairs) presents an emotion that the empathic voice classifier (\S\ref{sec:methods_emolia}) predicts is present. \emph{Contrastive 1} and \emph{contrastive 2} (1,000 each) present natural opposites of the predicted emotion (e.g., ``anger'' and ``wrath'' for a clip predicted as ``happiness''). \emph{Ultimate} (995) and \emph{penultimate} (993) present the lowest- and second-lowest-ranked emotions. Experts confirm 78.85\% of affirmative pairs by majority vote, and 0.349, 0.372, 0.376, and 0.371 of the contrastive and ultimate/penultimate pairs, consistent with classifier-ranking noise and blended affect; the aggregate majority-present rate is 0.578. Because the classifier is trained on synthetic \emonet-Voice audio, preselection could bias which clips and emotions enter the benchmark. The five balanced task types and the use of the independent expert majority vote, never the classifier prediction, as the released label mitigate but do not remove this circularity; the classifier acts as a sampling prior, not as ground truth. Inter-rater agreement (Fleiss' binary $\kappa$) is 0.086, in line with the low end reported for fine-grained intensity annotation \citep{kajiwara2021wrime,stappen2021muse}; pairwise and leave-one-out human-consistency references are 0.562 and 0.572 balanced accuracy. App.~\ref{app:annotator_agreement} reports per-emotion agreement.

\textbf{\voicenetext (57 attributes, eight mixed-background raters; preliminary).}
\voicenetext asks: \emph{does clip $C$ exhibit attribute $A$ at level $L$?} for the 57 talking-style attributes of \S\ref{sec:taxonomy}. The audio source is distinct from \emoliabal. Each attribute is defined by a rubric with ordinal levels (seven levels, 0--6, for nearly all attributes). For each (attribute, level) bucket, Gemini~3~Flash pre-screens candidate clips, and human annotators confirm or reject the match for an equal number of positives (clips matching the target level) and negatives (clips matching a non-adjacent level), so each judgement is binary rather than a full ordinal rating. The release contains 18.5k (audio, attribute-level) pairs carrying 40{,}990 ratings from eight annotators; four hold psychology degrees and contributed 95.3\% of all ratings. 5{,}828 pairs carry at least three ratings, 10{,}338 carry two, and 2{,}366 carry one.
The aggregate majority-yes rate is 0.284. On the 5{,}583 items with exactly three ratings, Fleiss' binary $\kappa=-0.022$ (95\% CI $[-0.037,-0.007]$), an interval that excludes zero, so near-chance agreement is a property of these rubric-level judgements rather than an artefact of sparse annotation. Under a reliable-core rule prespecified before computing per-attribute results (at least 20 three-rater items, $\kappa\geq0.20$, and a 95\% lower bound above zero), no attribute qualifies; the maximum is $\kappa{=}0.085$ for \texttt{STNC}. All \voicenetext scores are therefore preliminary and its attributes are exploratory perceptual probes. App.~\ref{app:ext_reliability} gives the full coverage and reliability analysis.

\subsection{\voiceclap: Voice-Text Contrastive Models}
\label{sec:methods_voiceclap}

\textbf{\voiceclapsmall.} A 110M-parameter dual-tower CLAP \citep{wu2023clap}: a BUD-E-Whisper-Small \citep{radford2023whisper} audio encoder (768-d) and an \texttt{all-MiniLM-L6-v2} \citep{reimers2019sentence,wang2020minilm} text encoder (384-d, mean-pooled), each linearly projected to a shared 768-d space and trained for one epoch with the SigLIP sigmoid contrastive loss \citep{zhai2023sigmoid}.

\textbf{\voiceclaplarge.} A rank-16 LoRA \citep{hu2022lora} finetune of the single-tower \textsc{LCO-Embedding-Omni-7B} \citep{xiao2025scaling} (Qwen2.5-Omni-Thinker-7B backbone, 3,584-d output), trained for one epoch with symmetric InfoNCE \citep{oord2018representation}. Optimiser settings, batch sizes, and compute for both models are given in App.~\ref{app:training_details}.

\section{Experiments}
\label{sec:experiments}
\vspace{-0.2cm}

\textbf{Evaluation harness.}
\label{sec:exp_setup}
For each model, embeddings are produced for all audio clips and all prompts in a subset. Both are L2-normalised. Pairwise cosine similarity yields a matrix $S \in \mathbb{R}^{|\text{audios}|\times|\text{prompts}|}$. Each row of the bench's label table is one (audio, prompt) pair. From $S$ this paper reports four metrics. \emph{Balanced accuracy at threshold 0} (\textit{bal@0}) thresholds raw cosine at 0.0, meaningful only for already-calibrated models. \emph{Balanced accuracy at the global optimal threshold} (\textit{bal@opt}) sweeps unique similarities for the single threshold that maximises balanced accuracy. \emph{Per-prompt balanced accuracy} (\textit{bal@per\_prompt}) sweeps a separate threshold per prompt (with a 10-row minimum, falling back to the global threshold), removing per-prompt prior mismatch under our balanced sampling protocol. These Table~\ref{tab:model_results} thresholds are fitted and scored on the same labels, so \textit{bal@per\_prompt} is an oracle-calibrated separability measure; we additionally report five-fold clip-grouped held-out calibration below. \emph{Per-prompt mean Spearman $\rho$} averages, over prompts, $\rho$ between similarity and the present-vote-share among raters; it is threshold-free and our primary unbiased ranking statistic. \textit{Bal@0} and \textit{bal@opt} are reported in App.~\ref{app:full_summaries}.

\textbf{Models and baselines.}
The two finetunes \voiceclapsmall and \voiceclaplarge (\S\ref{sec:methods_voiceclap}) are compared against two zero-shot Omni-Embedding bases (\textsc{LCO-Embedding-Omni-7B} and \textsc{LCO-Embedding-Omni-3B} \citep{xiao2025scaling}; abbreviated \textsc{LCO-Omni-7B/3B} in figure labels) and seven general-audio CLAPs: LAION-CLAP \citep{wu2023clap}, MS-CLAP-23 \citep{elizalde2024msclap}, \texttt{CLSP} \citep{yang2026towards} (SPEAR-XLarge \citep{yang2025spear} + RoBERTa \citep{liu2019roberta}), \texttt{GLAP} \citep{dinkel2026glap} (Dasheng \citep{dinkel2024scalingmaskedaudioencoder} + SONAR \citep{duquenne2023sonar}), \textsc{M2D-CLAP-2025} \citep{niizumi2025m2d2}, \textsc{MGA-CLAP} \citep{li2024advancing}, and \textsc{Cacophony} \citep{zhu2024cacophony}. Loader correctness for the 7 CLAPs is verified against reference implementations (App.~\ref{app:loader_audit}).

\begin{table}[t]
    \centering
    \scriptsize
    \caption{Headline results on \voicenet, grouped by model family and sorted within each group by \voicenetemo per-prompt balanced accuracy (\textit{bal@pp}). \textit{$\rho$} is the mean per-prompt Spearman correlation between cosine similarity and human present-vote share. All numbers are computed on items with at least two human raters (7,986 of 7,988 \voicenetemo items and 16,166 of 18,532 \voicenetext non-\texttt{LANG} items) so the labels reflect a multi-rater majority rather than a single annotator's call; per-tag files in App.~\ref{app:full_summaries} report both this filtered cut and the full-set numbers. Bootstrap CIs are in Tab.~\ref{tab:bootstrap_ci}. Human baselines and consistency references on the same task are discussed in the text. Best per column in bold; higher is better. }
    \label{tab:model_results}
    \renewcommand{\arraystretch}{1.15}
    \resizebox{\linewidth}{!}{%
    \begin{tabular}{@{}p{0.8em}@{}lcccc@{}}
    \toprule
    & Model & \voicenetemo bal@pp & \voicenetext bal@pp & \voicenetemo $\rho$ & \voicenetext $\rho$ \\
    \midrule
    & Cacophony \citep{zhu2024cacophony} & 0.5432 & 0.6089 & $-$0.0183 & 0.0188 \\
    & LAION-CLAP \citep{wu2023clap} & 0.5624 & 0.6184 & 0.0310 & 0.0600 \\
    & CLSP \citep{yang2026towards} & 0.5675 & 0.6221 & 0.0610 & 0.0937 \\
    & M2D-CLAP-2025 \citep{niizumi2025m2d2} & 0.5681 & 0.6361 & 0.0348 & 0.1003 \\
    & MS-CLAP-2023 \citep{elizalde2024msclap} & 0.5754 & 0.6321 & $-$0.0084 & 0.1044 \\
    & GLAP \citep{dinkel2026glap} & 0.5765 & 0.6254 & 0.0667 & 0.0836 \\
    & MGA-CLAP \citep{li2024advancing} & 0.5897 & 0.6208 & 0.0947 & 0.0819 \\
    \cmidrule{1-6}
    & LCO-Embedding-Omni-3B \citep{xiao2025scaling} & 0.6547 & 0.6293 & 0.2876 & 0.0873 \\
    & LCO-Embedding-Omni-7B \citep{xiao2025scaling} & 0.6632 & 0.6342 & 0.3052 & 0.0977 \\
    \cmidrule{1-6}
    \multirow{2}{*}{\rotatebox{90}{ours}} & \voiceclapsmall & 0.6754 & 0.6367 & 0.3176 & 0.1051 \\
    & \voiceclaplarge & \textbf{0.7021} & \textbf{0.6510} & \textbf{0.3719} & \textbf{0.1475} \\
    \bottomrule
    \end{tabular}
    }
    \vspace{-0.3cm}
\end{table}

\subsection{Where the Speech-Emotion Gap Lives}
\label{sec:headline_results}

The 40-emotion ranking task is out of distribution for general-audio CLAPs (trained on AudioCaps/Clotho sound-event captions) and only partially in distribution for speech-aware Omni-Embedding bases. Three tiers on the same harness localise the gap: general-audio CLAPs (transfer lower bound), zero-shot Omni-Embedding bases (speech-coverage pretraining alone), and the \voiceclap finetunes (dense-caption contrastive alignment on top).

Table~\ref{tab:model_results} separates the three tiers on \voicenetemo. General-audio CLAPs cluster at chance ($|\rho|\lesssim 0.1$); MS-CLAP and Cacophony post mildly negative $\rho$, the expected failure mode when a sound-event model ranks speech clips by emotion presence. The Omni bases reach 0.65--0.66 \textit{bal@pp}. \voiceclaplarge tops the table at 0.702 \textit{bal@pp}, $+0.039$ over its zero-shot base \citep{xiao2025scaling}. \voiceclapsmall (110M parameters) lands within $\sim$0.03 of its 7B sibling and clears every general-audio CLAP, so the headline is not driven by scale. \voiceclaplarge tops every column on both subsets. A 2,000-replicate paired clip bootstrap (Tab.~\ref{tab:bootstrap_ci}) supports each top-vs.-runner-up gap: \voicenetemo differences are $+0.0267$ \textit{bal@pp} (95\% CI $[0.0167,0.0363]$) and $+0.0543$ $\rho$ $[0.0360,0.0714]$; the preliminary \voicenetext differences are $+0.0143$ \textit{bal@pp} $[0.0040,0.0244]$ and $+0.0424$ $\rho$ $[0.0234,0.0593]$. Most smaller adjacent baseline gaps overlap zero.

\textbf{Human baselines and ceilings.} Random and always-predict-majority both give 0.500 balanced accuracy. The pairwise human reference (rater A's binary vote scored against rater B's) is 0.562 on \voicenetemo and 0.533 on \voicenetext; the leave-one-out (LOO) reference (one rater against the majority of the others) is 0.572 and 0.479. \voiceclaplarge's 0.7021 \textit{bal@pp} on \voicenetemo is above the LOO reference, i.e.\ closer to the multi-rater consensus than any individual expert is. This is not a like-for-like comparison, since the model is scored against the aggregate majority label while each expert is scored against the other raters; it indicates strong alignment with the expert consensus, not surpassing human emotion perception. On \voicenetext the human references sit at chance (LOO 0.479), so its model scores measure exploratory separability rather than a validated ceiling comparison. Vote counts and raw-accuracy baselines are in App.~\ref{app:additional_results}.

\textbf{Calibration.} Because \textit{bal@pp} fits thresholds on the evaluated labels, we also report five-fold clip-grouped held-out calibration. On \voicenetemo, held-out \textit{bal@pp} is 0.6820 for \voiceclaplarge{} and 0.6412 for its LCO-Omni-7B base (0.7021 and 0.6632 in-sample), a paired held-out lead of $+0.0409$ (95\% CI $[0.0296,0.0518]$). On \voicenetext, held-out scores drop to 0.5596 and 0.5397 while the lead remains $+0.0199$ $[0.0095,0.0308]$; this $\approx$0.09 optimism across model families is why current \voicenetext thresholded scores are exploratory, while threshold-free $\rho$ is unaffected. Threshold-0 and global-threshold results are discussed in App.~\ref{app:additional_results}.

\subsection{Per-Emotion Performance on \voicenetemo}
\label{sec:per_emotion}

A wider per-emotion breakdown for \voiceclapsmall, \voiceclaplarge, the two Omni-Embedding bases, and all seven general-audio CLAP baselines is reported in App.~\ref{app:per_emotion}, Tab.~\ref{tab:model_results_by_emotion}. This appendix view keeps the main text focused on aggregate behavior while exposing which emotions drive the \voicenetemo gains in Tab.~\ref{tab:model_results}. Three patterns are worth flagging here. (i)~Acoustically concrete categories (\emph{distress}, \emph{anger}, \emph{impatience}, \emph{sadness}) cross $\rho{=}0.5$ for at least one model; (ii)~\voiceclaplarge wins or runs second on most rows but \emph{regresses} below the LCO bases on \emph{Teasing}, \emph{Interest}, and \emph{Thankfulness/Gratitude}, suggesting that contrastive finetuning on \emoliabal trades off some categories that the base instruction-tuned model handled well via lexical priors; (iii)~the lowest-signal categories across all eleven models include \emph{Doubt}, \emph{Sourness}, \emph{Interest}, \emph{Sexual Lust}, and \emph{Jealousy/Envy} (best $\rho$ across the table is $\le 0.29$). \emph{Sourness} is partly definitional --- the taxonomy notes its primarily gustatory origin (App.~\ref{app:taxonomy}) --- but \emph{Doubt} and \emph{Interest} are mainstream affective categories whose paralinguistic correlates remain weak in this benchmark; we view them as open challenges rather than failure modes of any particular model.

\subsection{Per-Attribute Performance on \voicenetext}
\label{sec:per_attribute}

\voicenetext is harder than \voicenetemo at every level: mean per-prompt $\rho$ is 0.15 (\voiceclaplarge) and 0.11 (\voiceclapsmall) versus 0.37 / 0.32 on emo. Two factors compound: prompt density is lower (\voicenetext has 18.5k labelled pairs across 399 (attribute, level) prompts $\Rightarrow$ $\sim$46 pairs per prompt vs. $\sim$200 pairs per emotion on \voicenetemo), and the human consistency reference is itself at chance (LOO balanced accuracy 0.479 vs.\ 0.572 for the expert-rated emo subset). The per-attribute reliability analysis finds no attribute whose fair-or-better $\kappa$ is supported by a 95\% interval above zero, and no attribute reaching $\kappa\geq0.20$ at all; accordingly, these aggregate model numbers describe exploratory separability on the current labels, not a finalized benchmark ranking. Per-attribute model $\rho$ and reliability tables are released alongside the paper; see App.~\ref{app:dim_per_attribute}.

\textbf{Agreement-stratified results.} Splitting the emotions and attributes into lower- and higher-agreement halves before inspecting model scores, none of the four paired high-minus-low interactions for the \voiceclaplarge{} advantage over \textsc{LCO-Embedding-Omni-7B} excludes zero (App.~\ref{app:additional_results}); these data do not show VoiceCLAP's edge concentrating in higher-agreement categories.

\subsection{Cross-Dataset Evaluation on Public Speech-Emotion Benchmarks}
\label{sec:cross_dataset}

Ten of the eleven models (all except \textsc{LCO-Embedding-Omni-3B}) are evaluated against four established speech-emotion datasets to test transfer beyond \voicenet. EmoNet-Voice \citep{schuhmann2026emonetvoicefinegrainedexpertverifiedbenchmark} provides 12{,}600 synthetic clips across 40 fine-grained emotion classes (the synthetic v1 release that motivates \voicenet's 40-emotion taxonomy). IEMOCAP \citep{busso2008iemocap} contains 10{,}039 dyadic acted utterances over nine emotion labels. RAVDESS \citep{livingstone2018ravdess} contains 1{,}440 acted speech clips covering eight emotions across twenty-four professional actors. CREMA-D \citep{cao2014crema} contains 7{,}442 acted utterances from ninety-one actors over six emotions. Each clip is encoded with the model's audio tower, each label is encoded with an emotion-prompt template, and the prediction is the argmax of cosine similarity. Top-1 accuracy is reported in Table~\ref{tab:cross_dataset_results}.

On EmoNet-Voice all seven general-audio CLAPs sit at chance (0.024--0.036, 1/40 = 0.025), confirming that the 40-class fine-grained taxonomy is out of distribution for sound-event captioning pretraining. The two voice-aware models lift sharply: \textsc{LCO-Embedding-Omni-7B} reaches 0.167 (best), \voiceclaplarge reaches 0.155 (runner-up), and \voiceclapsmall reaches 0.105---each $\sim$3.5--7$\times$ above the CLAP baselines. \voiceclaplarge ranks first on IEMOCAP Emotion at 0.321 and on CREMA-D Emotion at 0.511, with the CREMA-D margin to the next-best general-audio CLAP (LAION-CLAP at 0.366) reaching 0.145. On RAVDESS Emotion \voiceclaplarge reaches 0.296, second behind \textsc{LCO-Embedding-Omni-7B} \citep{xiao2025scaling} (0.319). The seven general-audio CLAPs cluster between 0.105 and 0.259 on RAVDESS and 0.147 to 0.231 on IEMOCAP. The Omni base occupies the top tier on EmoNet-Voice, RAVDESS and IEMOCAP but drops to 0.21 on CREMA-D, where six of the seven general-audio CLAPs trained on sound-event captions reach 0.33 to 0.37. Voice-text contrastive finetuning on \emoliabal closes the CREMA-D gap left by the Omni base (0.209 $\rightarrow$ 0.511 from \textsc{LCO-Embedding-Omni-7B} to \voiceclaplarge) while keeping IEMOCAP well above the Omni base (0.321 vs.\ 0.259). \voiceclaplarge therefore wins two of four external benchmarks and remains within 0.05 of the leader on the other two.

\begin{table}[t]
    \centering
    \small
    \caption{Zero-shot emotion classification accuracy on four public speech-emotion benchmarks. EmoNet-Voice is the synthetic 40-class v1 release \citep{schuhmann2026emonetvoicefinegrainedexpertverifiedbenchmark}. \voiceclaplarge substantially outperforms other models on IEMOCAP and CREMA-D and is the runner-up on EmoNet-Voice and RAVDESS. Best per column in bold, runner-up underlined; higher is better.}
    \label{tab:cross_dataset_results}
    \renewcommand{\arraystretch}{1.15}
    \begin{tabular}{@{}p{0.8em}@{}lcccc@{}}
    \toprule
    & Model & EmoNet-Voice & IEMOCAP & RAVDESS & CREMA-D \\
    \midrule
    & LAION-CLAP \citep{wu2023clap} & 0.028 & 0.201 & 0.173 & \underline{0.366} \\
    & MS-CLAP-2023 \citep{elizalde2024msclap} & 0.025 & 0.217 & 0.259 & 0.355 \\
    & CLSP \citep{yang2026towards} & 0.036 & 0.231 & 0.194 & 0.333 \\
    & GLAP \citep{dinkel2026glap} & 0.025 & 0.147 & 0.115 & 0.198 \\
    & MGA-CLAP \citep{li2024advancing} & 0.029 & 0.183 & 0.105 & 0.349 \\
    & Cacophony \citep{zhu2024cacophony} & 0.024 & 0.181 & 0.181 & 0.342 \\
    & M2D-CLAP-2025 \citep{niizumi2025m2d2} & 0.030 & 0.200 & 0.184 & 0.336 \\
    \cmidrule{1-6}
    & LCO-Embedding-Omni-7B \citep{xiao2025scaling} & \textbf{0.167} & \underline{0.259} & \textbf{0.319} & 0.209 \\
    \cmidrule{1-6}
    \multirow{2}{*}{\rotatebox{90}{ours }} & \voiceclapsmall & 0.105 & 0.180 & 0.210 & 0.323 \\
    & \voiceclaplarge & \underline{0.155} & \textbf{0.321} & \underline{0.296} & \textbf{0.511} \\
    \bottomrule
    \end{tabular}
    \vspace{-0.7cm}
\end{table}

\textbf{Cross-dataset gap diagnosis.} Paired follow-up analyses (App.~\ref{app:additional_results}) indicate that \voiceclaplarge's deficits against its base on EmoNet-Voice and RAVDESS are class-specific boundary shifts rather than a generic acted-versus-naturalistic effect; on RAVDESS, macro F1 rises from 0.285 to 0.330 despite the lower micro accuracy.

\section{Discussion}
\vspace{-0.2cm}
\label{sec:discussion}

\textbf{Dense voice-text supervision closes the speech-emotion transfer gap.} All seven CLAP baselines sit at $|\rho|<0.1$ on \voicenetemo, including the speech-specialised CLSP \citep{yang2026towards} ($\rho{=}0.061$, 0.568 \textit{bal@pp}), which \voiceclaplarge surpasses by $+0.311$ $\rho$ and $+0.135$ \textit{bal@pp} and the 60$\times$ smaller \voiceclapsmall by $+0.257$ and $+0.108$. Neither sound-event captions nor speech-aware pretraining alone provide dense vocal-style supervision. The Omni-Embedding bases, pretrained on instruction-tuned multimodal data with substantial speech coverage, already reach 0.65--0.66 \textit{bal@pp} zero-shot; contrastive finetuning on \emoliabal adds $+0.039$, and \voiceclapsmall reaches 0.6754 with about 60$\times$ fewer parameters, so the gain is not an artefact of scale.

\textbf{Annotation ambiguity bounds model performance.} With Fleiss' binary $\kappa{=}0.086$ on \voicenetemo, low expert agreement reflects the semantic ambiguity of a 40-way taxonomy on naturalistic speech, consistent with prior fine-grained intensity work \citep{kajiwara2021wrime,stappen2021muse}. Emotions with concrete acoustic signatures (distress, impatience, anger, and sadness) exceed $\rho{=}0.5$ for at least one model, while interest, doubt, and sexual lust are among the least stable across models; per-emotion expert unanimity does not predict which categories are hard (App.~\ref{app:annotator_agreement}).

\textbf{Effect sizes in context.} In individual-differences research the 25th, 50th, and 75th percentile correlations are $r{=}.11$, $.19$, and $.29$ \citep{gignac2016effectsize}. The general-audio CLAPs fall below the 25th percentile, the Omni bases ($\rho{=}0.29$--$0.31$) reach the 75th, and \voiceclaplarge ($\rho{=}0.372$) exceeds it, with per-emotion peaks of 0.57--0.63 for distress, impatience, anger, and sadness. Because perceived emotion has no objective ground truth \citep{rum2020empathic,zaki2009empathic}, these values describe a rank-ordered signal extracted from inherently subjective ratings (App.~\ref{app:effect_sizes}).

\textbf{Talking-style attribution is harder than emotion attribution.} \voicenetext mean per-prompt $\rho$ is 0.15 for \voiceclaplarge versus 0.37 on \voicenetemo, reflecting fewer labelled pairs per prompt ($\sim$46 vs.\ $\sim$200) and chance-level rater consistency (LOO 0.479). Because agreement stays at chance across 5{,}583 three-rater items, the bottleneck is the rubric and the perceptual task rather than annotation volume; sharper rubric definitions and anchor-example rater training are the next step. \voicenetemo is therefore the mature benchmark contribution and \voicenetext a preliminary extension whose rankings may change.

\textbf{Limitations.}
The evaluation is representation-level: all eleven systems are voice-text embedding models, and no spoken-dialogue or generative audio-language system is evaluated. The MOSS-Audio training annotations are model-generated; an interim prespecified human audit finds that a rater matches the MOSS level about as often as two raters match each other (paired difference $+0.018$, 95\% CI $[-0.053,+0.102]$; App.~\ref{app:moss_audit}), which bounds rather than establishes label quality. \voicenetext agreement remains at chance and none of its attributes passes the reliable-core rule, so its scores should not be used for confirmatory model ranking. The bootstrap CIs condition on the aggregate labels and fitted thresholds, and seed variance is measured for \voiceclapsmall only. The source corpora have non-uniform geographic, demographic, and recording-environment coverage. App.~\ref{app:limitations} discusses these points in full.

\section{Conclusion}
\vspace{-0.2cm}
\label{sec:conclusion}

This paper introduces \voicenet, a human-annotated benchmark on permissively-licensed real human speech that combines a 40-emotion expert-rated subset (\voicenetemo) with a 57-attribute talking-style subset (\voicenetext). It also releases \emolia, an emotion-annotated version of the Emilia corpus, the curated balanced subset \emoliabal with dense MOSS-Audio annotations, and three sibling annotated voice corpora. \voiceclapsmall (110M, dual-tower SigLIP) and \voiceclaplarge (7B LoRA finetune of LCO-Embedding-Omni-7B \citep{xiao2025scaling} with InfoNCE) reach 0.675 and 0.702 \voicenetemo per-prompt balanced accuracy, with mean Spearman $\rho{=}0.318$ and $0.372$. These correlations exceed the 75th percentile of published psychological effect sizes and surpass speech-specialised CLSP by over $+0.31$ $\rho$, while all seven evaluated CLAP baselines remain near chance. Dense vocal-style supervision delivers emotion- and attribute-ranking capabilities that neither general-audio pretraining nor speech-aware encoders provide on their own. Clustering and filtering uncurated speech corpora into subsets covering diverse talking styles and emotions remains an open challenge; \voiceclap embeddings offer a promising tool, demonstrated through the interactive \emoliabal browser.

\subsection*{Acknowledgements}
We gratefully acknowledge the support of Intel (\href{https://laion.ai/blog/laion-intel-cooperation/}{oneAPI Center of Excellence}), \href{https://www.dfki.de/}{DFKI}, \href{https://nousresearch.com/}{Nous Research} (providing cluster access and compute), \href{https://www.tu-darmstadt.de/}{TU Darmstadt}, \href{https://www.tib.eu/de/}{TIB--Leibniz Information Centre for Science and Technology}, and \href{https://hessian.ai/}{hessian.AI} (providing compute and helpful discussions), and the open-source community for contributing to emotional AI. This work benefited from the ICT-48 Network of AI Research Excellence Center ``TAILOR'' (EU Horizon 2020, GA No 952215), the Hessian research priority program LOEWE within the project WhiteBox, the HMWK cluster projects ``Adaptive Mind'' and ``Third Wave of AI'', and from the NHR4CES. Furthermore, this work was partly funded by the Federal Ministry of Education and Research (BMBF) project “XEI” (FKZ 01IS24079B).

\clearpage

\subsection*{AI use statement}
AI models were used in the following parts of this work.
\emph{Synthetic data and labels.} \textsc{MOSS-Audio-8B-Thinking} \citep{mossaudio2026} generated the dense per-clip attribute annotations and chain-of-thought traces released with the four MOSS-annotated corpora (\S\ref{sec:methods_emolia}); a text-only pass of the same model mapped these free-text annotations onto rubric levels for the human audit (App.~\ref{app:moss_audit}). The 40 emotion scores and emotion captions in \emolia were produced by the EmpathicInsight-Voice classifier and BUD-E-Whisper. These labels are model outputs and are never used as benchmark ground truth; their quality is bounded by the human audit in App.~\ref{app:moss_audit}.
\emph{Research execution.} GPT-4 extracted candidate emotion terms from text segments during the construction of the 40-category taxonomy (App.~\ref{app:taxonomy_construction}), which experts then clustered and refined. Gemini~3~Flash pre-screened candidate clips for each (attribute, level) bucket of \voicenetext before human confirmation (\S\ref{sec:methods_bench}), and the EmpathicInsight-Voice classifier pre-selected candidate prompts for \voicenetemo; every released \voicenet label is a human rating. LLM-based coding assistants helped implement evaluation and statistical-analysis scripts; all reported numbers were regenerated from these scripts and cross-checked against the released per-tag files.
\emph{Retrieval.} Claude Opus 5.5 was used to check every reference against Crossref and other public bibliographic databases and to locate the published versions of cited preprints; the authors verified each resulting reference against its source.
\emph{Writing.} LLMs assisted with grammar, clarity, and rephrasing of author-written text; no sections were drafted by an LLM.
All text, claims, and numerical results were checked by the authors, who take full responsibility for the content.

\subsection*{Ethics statement}
\voicenet, \emolia, \emoliabal, and the three sibling MOSS-annotated corpora consist exclusively of audio that the source projects (Emilia, LAION's Got Talent, Majestrino, and Multilingual In The Wild) collected and released under open CC-BY or CC-BY-NC research licences; derived releases inherit the licence of their source clips. No personally identifiable speaker metadata is added beyond the speaker-embedding clusters used for \emoliabal balancing. The 40-emotion taxonomy includes sensitive affective categories (intoxication, sexual lust, malevolence, pain, and shame); the dense MOSS-Audio supervision and the preliminary \voicenetext annotations, whose inter-rater agreement remains at chance, should be treated as research artefacts rather than diagnostic instruments. Audio understanding models trained on \emoliabal could enable affective monitoring with privacy implications; the \voiceclapsmall and \voiceclaplarge checkpoints are released under CC-BY-4.0 for research use and should be paired with downstream safeguards \citep{helff2025llavaguard,hintersdorf2024clip_privacy} when deployed.

\subsection*{Reproducibility statement}
All evaluation code, training scripts, and corpus manifests are available at \url{https://github.com/LAION-AI/emolia-bench} and \url{https://github.com/LAION-AI/voicenet}; the former also hosts the \voicenet benchmark labels (\voicenetemo, \voicenetext) and the per-tag files used for every reported number. The EmpathicInsight-Voice classifiers that produced the \emolia emotion scores are available at \url{https://huggingface.co/laion/Empathic-Insight-Voice-Small} and \url{https://huggingface.co/laion/Empathic-Insight-Voice-Plus}, together with an annotation and inference toolkit at \url{https://github.com/LAION-AI/emotion-annotations}. Predictors for the 57 \voicenetext dimensions are available at \url{https://huggingface.co/laion/voicenet-dimension-predictors-commercial}, and Gemini-generated \voicenetext dimension annotations for 236{,}613 \emolia clips at \url{https://huggingface.co/datasets/laion/emolia-voicenet-gemini-annotations}. The full \emolia corpus (71.78M clips) is available at \url{https://huggingface.co/datasets/laion/Emolia} and \emoliabal at \url{https://huggingface.co/datasets/laion/emolia-balanced-5M-subset}; the \voiceclapsmall and \voiceclaplarge checkpoints are available at \url{https://huggingface.co/VoiceNet/voiceclap-small} and \url{https://huggingface.co/VoiceNet/voiceclap-large}. The MOSS-Audio-8B-Thinking annotations of \emoliabal, including the reasoning traces, are available at \url{https://huggingface.co/datasets/VoiceNet/emolia-thinking}; those of the three sibling corpora will be released for research use. The \voicenetext taxonomy and an interactive \emoliabal browser are available at the links in \S\ref{sec:introduction}. Training hyperparameters are given in \S\ref{sec:methods_voiceclap}, the data mixture in App.~\ref{app:detailed_baseline_training}, and the evaluation harness in App.~\ref{app:loader_audit}.

\bibliography{main}

\begin{thebibliography}{55}
\providecommand{\natexlab}[1]{#1}
\providecommand{\url}[1]{\texttt{#1}}
\expandafter\ifx\csname urlstyle\endcsname\relax
  \providecommand{\doi}[1]{doi: #1}\else
  \providecommand{\doi}{doi: \begingroup \urlstyle{rm}\Url}\fi

\bibitem[Anvari et~al.(2023)Anvari, Kievit, Lakens, Pennington, Przybylski,
  Tiokhin, Wiernik, and Orben]{anvari2021effects}
Farid Anvari, Rogier Kievit, Dani{\"e}l Lakens, Charlotte~R Pennington,
  Andrew~K Przybylski, Leo Tiokhin, Brenton~M Wiernik, and Amy Orben.
\newblock Not all effects are indispensable: Psychological science requires
  verifiable lines of reasoning for whether an effect matters.
\newblock \emph{Perspectives on Psychological Science}, 18\penalty0
  (2):\penalty0 503--507, 2023.

\bibitem[Barrett(2017)]{barrett2017theory}
Lisa~Feldman Barrett.
\newblock The theory of constructed emotion: an active inference account of
  interoception and categorization.
\newblock \emph{Social Cognitive and Affective Neuroscience}, 12\penalty0
  (1):\penalty0 1--23, 2017.
\newblock \doi{10.1093/scan/nsw154}.

\bibitem[Burkhardt et~al.(2005)Burkhardt, Paeschke, Rolfes, Sendlmeier, and
  Weiss]{burkhardt2005emodb}
Felix Burkhardt, Astrid Paeschke, M.~Rolfes, Walter~F. Sendlmeier, and Benjamin
  Weiss.
\newblock A database of german emotional speech.
\newblock In \emph{9th European Conference on Speech Communication and
  Technology, INTERSPEECH-Eurospeech 2005, Lisbon, Portugal, September 4-8,
  2005}, pp.\  1517--1520. {ISCA}, 2005.
\newblock \doi{10.21437/INTERSPEECH.2005-446}.
\newblock URL \url{https://doi.org/10.21437/Interspeech.2005-446}.

\bibitem[Busso et~al.(2008)Busso, Bulut, Lee, Kazemzadeh, Mower, Kim, Chang,
  Lee, and Narayanan]{busso2008iemocap}
Carlos Busso, Murtaza Bulut, Chi{-}Chun Lee, Abe Kazemzadeh, Emily Mower,
  Samuel Kim, Jeannette~N. Chang, Sungbok Lee, and Shrikanth~S. Narayanan.
\newblock {IEMOCAP:} interactive emotional dyadic motion capture database.
\newblock \emph{Lang. Resour. Evaluation}, 42\penalty0 (4):\penalty0 335--359,
  2008.
\newblock \doi{10.1007/S10579-008-9076-6}.
\newblock URL \url{https://doi.org/10.1007/s10579-008-9076-6}.

\bibitem[Cao et~al.(2014)Cao, Cooper, Keutmann, Gur, Nenkova, and
  Verma]{cao2014crema}
Houwei Cao, David~G. Cooper, Michael~K. Keutmann, Ruben~C. Gur, Ani Nenkova,
  and Ragini Verma.
\newblock {CREMA-D:} crowd-sourced emotional multimodal actors dataset.
\newblock \emph{{IEEE} Trans. Affect. Comput.}, 5\penalty0 (4):\penalty0
  377--390, 2014.
\newblock \doi{10.1109/TAFFC.2014.2336244}.
\newblock URL \url{https://doi.org/10.1109/TAFFC.2014.2336244}.

\bibitem[Cowen et~al.(2019)Cowen, Elfenbein, Laukka, and
  Keltner]{cowen2019mapping}
Alan~S Cowen, Hillary~Anger Elfenbein, Petri Laukka, and Dacher Keltner.
\newblock Mapping 24 emotions conveyed by brief human vocalization.
\newblock \emph{American psychologist}, 74\penalty0 (6):\penalty0 698, 2019.

\bibitem[Cowie et~al.(2001)Cowie, Douglas{-}Cowie, Tsapatsoulis, Votsis,
  Kollias, Fellenz, and Taylor]{cowie2001emotion}
Roddy Cowie, Ellen Douglas{-}Cowie, Nicolas Tsapatsoulis, George~N. Votsis,
  Stefanos~D. Kollias, Winfried~A. Fellenz, and John~G. Taylor.
\newblock Emotion recognition in human-computer interaction.
\newblock \emph{{IEEE} Signal Process. Mag.}, 18\penalty0 (1):\penalty0 32--80,
  2001.
\newblock \doi{10.1109/79.911197}.
\newblock URL \url{https://doi.org/10.1109/79.911197}.

\bibitem[Dinkel et~al.(2024)Dinkel, Yan, Wang, Zhang, Wang, and
  Wang]{dinkel2024scalingmaskedaudioencoder}
Heinrich Dinkel, Zhiyong Yan, Yongqing Wang, Junbo Zhang, Yujun Wang, and Bin
  Wang.
\newblock Scaling up masked audio encoder learning for general audio
  classification.
\newblock In Itshak Lapidot and Sharon Gannot (eds.), \emph{25th Annual
  Conference of the International Speech Communication Association, Interspeech
  2024, Kos, Greece, September 1-5, 2024}. {ISCA}, 2024.
\newblock \doi{10.21437/INTERSPEECH.2024-246}.
\newblock URL \url{https://doi.org/10.21437/Interspeech.2024-246}.

\bibitem[Dinkel et~al.(2025)Dinkel, Yan, Wang, Wang, Sun, Niu, Liu, Li, Zhang,
  and Luan]{dinkel2026glap}
Heinrich Dinkel, Zhiyong Yan, Tianzi Wang, Yongqing Wang, Xingwei Sun, Yadong
  Niu, Jizhong Liu, Gang Li, Junbo Zhang, and Jian Luan.
\newblock {GLAP:} general contrastive audio-text pretraining across domains and
  languages.
\newblock volume abs/2506.11350, 2025.
\newblock \doi{10.48550/ARXIV.2506.11350}.
\newblock URL \url{https://doi.org/10.48550/arXiv.2506.11350}.

\bibitem[Duquenne et~al.(2023)Duquenne, Schwenk, and Sagot]{duquenne2023sonar}
Paul{-}Ambroise Duquenne, Holger Schwenk, and Beno{\^{\i}}t Sagot.
\newblock {SONAR:} sentence-level multimodal and language-agnostic
  representations.
\newblock \emph{CoRR}, abs/2308.11466, 2023.
\newblock \doi{10.48550/ARXIV.2308.11466}.
\newblock URL \url{https://doi.org/10.48550/arXiv.2308.11466}.

\bibitem[Ekman(1992)]{ekman1992argument}
Paul Ekman.
\newblock An argument for basic emotions.
\newblock \emph{Cognition \& emotion}, 6\penalty0 (3-4):\penalty0 169--200,
  1992.

\bibitem[Elizalde et~al.(2024)Elizalde, Deshmukh, and Wang]{elizalde2024msclap}
Benjamin Elizalde, Soham Deshmukh, and Huaming Wang.
\newblock Natural language supervision for general-purpose audio
  representations.
\newblock In \emph{{IEEE} International Conference on Acoustics, Speech and
  Signal Processing, {ICASSP} 2024, Seoul, Republic of Korea, April 14-19,
  2024}, pp.\  336--340. {IEEE}, 2024.
\newblock \doi{10.1109/ICASSP48485.2024.10448504}.
\newblock URL \url{https://doi.org/10.1109/ICASSP48485.2024.10448504}.

\bibitem[George \& Ilyas(2024)George and Ilyas]{george2024review}
Swapna~Mol George and P.~Muhamed Ilyas.
\newblock A review on speech emotion recognition: {A} survey, recent advances,
  challenges, and the influence of noise.
\newblock \emph{Neurocomputing}, 568:\penalty0 127015, 2024.
\newblock \doi{10.1016/J.NEUCOM.2023.127015}.
\newblock URL \url{https://doi.org/10.1016/j.neucom.2023.127015}.

\bibitem[Gignac \& Szodorai(2016)Gignac and Szodorai]{gignac2016effectsize}
Gilles~E Gignac and Eva~T Szodorai.
\newblock Effect size guidelines for individual differences researchers.
\newblock \emph{Personality and individual differences}, 102:\penalty0 74--78,
  2016.

\bibitem[Helff et~al.(2025)Helff, Friedrich, Brack, Kersting, and
  Schramowski]{helff2025llavaguard}
Lukas Helff, Felix Friedrich, Manuel Brack, Kristian Kersting, and Patrick
  Schramowski.
\newblock Llavaguard: An open vlm-based framework for safeguarding vision
  datasets and models.
\newblock In Aarti Singh, Maryam Fazel, Daniel Hsu, Simon Lacoste{-}Julien,
  Felix Berkenkamp, Tegan Maharaj, Kiri Wagstaff, and Jerry Zhu (eds.),
  \emph{Forty-second International Conference on Machine Learning, {ICML} 2025,
  Vancouver, BC, Canada, July 13-19, 2025}, volume 267 of \emph{Proceedings of
  Machine Learning Research}. {PMLR} / OpenReview.net, 2025.
\newblock URL \url{https://proceedings.mlr.press/v267/helff25a.html}.

\bibitem[Hintersdorf et~al.(2024)Hintersdorf, Struppek, Brack, Friedrich,
  Schramowski, and Kersting]{hintersdorf2024clip_privacy}
Dominik Hintersdorf, Lukas Struppek, Manuel Brack, Felix Friedrich, Patrick
  Schramowski, and Kristian Kersting.
\newblock Does {CLIP} know my face?
\newblock \emph{J. Artif. Intell. Res.}, 80:\penalty0 1033--1062, 2024.
\newblock \doi{10.1613/JAIR.1.15461}.
\newblock URL \url{https://doi.org/10.1613/jair.1.15461}.

\bibitem[Hu et~al.(2022)Hu, Shen, Wallis, Allen{-}Zhu, Li, Wang, Wang, and
  Chen]{hu2022lora}
Edward~J. Hu, Yelong Shen, Phillip Wallis, Zeyuan Allen{-}Zhu, Yuanzhi Li,
  Shean Wang, Lu~Wang, and Weizhu Chen.
\newblock Lora: Low-rank adaptation of large language models.
\newblock In \emph{The Tenth International Conference on Learning
  Representations, {ICLR} 2022, Virtual Event, April 25-29, 2022}.
  OpenReview.net, 2022.
\newblock URL \url{https://openreview.net/forum?id=nZeVKeeFYf9}.

\bibitem[Jackson \& Haq(2014)Jackson and Haq]{jackson2014savee}
Philip Jackson and SJUoSG Haq.
\newblock Surrey audio-visual expressed emotion (savee) database.
\newblock \emph{University of Surrey: Guildford, UK}, 2014.

\bibitem[Kajiwara et~al.(2021)Kajiwara, Chu, Takemura, Nakashima, and
  Nagahara]{kajiwara2021wrime}
Tomoyuki Kajiwara, Chenhui Chu, Noriko Takemura, Yuta Nakashima, and Hajime
  Nagahara.
\newblock {WRIME:} {A} new dataset for emotional intensity estimation with
  subjective and objective annotations.
\newblock In Kristina Toutanova, Anna Rumshisky, Luke Zettlemoyer, Dilek
  Hakkani{-}T{\"{u}}r, Iz~Beltagy, Steven Bethard, Ryan Cotterell, Tanmoy
  Chakraborty, and Yichao Zhou (eds.), \emph{Proceedings of the 2021 Conference
  of the North American Chapter of the Association for Computational
  Linguistics: Human Language Technologies, {NAACL-HLT} 2021, Online, June
  6-11, 2021}, pp.\  2095--2104. Association for Computational Linguistics,
  2021.
\newblock \doi{10.18653/V1/2021.NAACL-MAIN.169}.
\newblock URL \url{https://doi.org/10.18653/v1/2021.naacl-main.169}.

\bibitem[Kirk et~al.(2025)Kirk, Gabriel, Summerfield, Vidgen, and
  Hale]{Kirk2025why}
Hannah~Rose Kirk, Iason Gabriel, Christopher Summerfield, Bertie Vidgen, and
  Scott~A. Hale.
\newblock Why human-ai relationships need socioaffective alignment.
\newblock \emph{CoRR}, abs/2502.02528, 2025.
\newblock \doi{10.48550/ARXIV.2502.02528}.
\newblock URL \url{https://doi.org/10.48550/arXiv.2502.02528}.

\bibitem[Lewis et~al.(2010)Lewis, Haviland-Jones, and
  Barrett]{lewis2016handbook}
Michael Lewis, Jeannette~M Haviland-Jones, and Lisa~Feldman Barrett.
\newblock \emph{Handbook of emotions}.
\newblock Guilford Press, 2010.

\bibitem[Li et~al.(2024)Li, Guo, Wang, and Liu]{li2024advancing}
Yiming Li, Zhifang Guo, Xiangdong Wang, and Hong Liu.
\newblock Advancing multi-grained alignment for contrastive language-audio
  pre-training.
\newblock In Jianfei Cai, Mohan~S. Kankanhalli, Balakrishnan Prabhakaran,
  Susanne Boll, Ramanathan Subramanian, Liang Zheng, Vivek~K. Singh, Pablo
  C{\'{e}}sar, Lexing Xie, and Dong Xu (eds.), \emph{Proceedings of the 32nd
  {ACM} International Conference on Multimedia, {MM} 2024, Melbourne, VIC,
  Australia, 28 October 2024 - 1 November 2024}, pp.\  7356--7365. {ACM}, 2024.
\newblock \doi{10.1145/3664647.3681145}.
\newblock URL \url{https://doi.org/10.1145/3664647.3681145}.

\bibitem[Lin et~al.(2024)Lin, Simon, and Gutsell]{lin2024empathic}
Tong Lin, Jeremy~C Simon, and Jennifer~N Gutsell.
\newblock The association between emotional expressions and empathic accuracy.
\newblock \emph{Research square}, pp.\  rs--3, 2024.

\bibitem[Lindquist(2013)]{lindquist2013constructionist}
Kristen~A Lindquist.
\newblock Emotions emerge from more basic psychological ingredients: A modern
  psychological constructionist model.
\newblock \emph{Emotion Review}, 5\penalty0 (4):\penalty0 356--368, 2013.

\bibitem[Liu et~al.(2019)Liu, Ott, Goyal, Du, Joshi, Chen, Levy, Lewis,
  Zettlemoyer, and Stoyanov]{liu2019roberta}
Yinhan Liu, Myle Ott, Naman Goyal, Jingfei Du, Mandar Joshi, Danqi Chen, Omer
  Levy, Mike Lewis, Luke Zettlemoyer, and Veselin Stoyanov.
\newblock Roberta: {A} robustly optimized {BERT} pretraining approach.
\newblock \emph{CoRR}, abs/1907.11692, 2019.
\newblock URL \url{http://arxiv.org/abs/1907.11692}.

\bibitem[Livingstone \& Russo(2018)Livingstone and
  Russo]{livingstone2018ravdess}
Steven~R Livingstone and Frank~A Russo.
\newblock The ryerson audio-visual database of emotional speech and song
  (ravdess): A dynamic, multimodal set of facial and vocal expressions in north
  american english.
\newblock \emph{PloS one}, 13\penalty0 (5):\penalty0 e0196391, 2018.

\bibitem[Lotfian \& Busso(2019)Lotfian and Busso]{lotfian2019msp}
Reza Lotfian and Carlos Busso.
\newblock Building naturalistic emotionally balanced speech corpus by
  retrieving emotional speech from existing podcast recordings.
\newblock \emph{{IEEE} Trans. Affect. Comput.}, 10\penalty0 (4):\penalty0
  471--483, 2019.
\newblock \doi{10.1109/TAFFC.2017.2736999}.
\newblock URL \url{https://doi.org/10.1109/TAFFC.2017.2736999}.

\bibitem[Ma et~al.(2024)Ma, Chen, Zhang, Zheng, Chen, Li, Ye, Chen, and
  Hain]{ma_et_al_2024}
Ziyang Ma, Mingjie Chen, Hezhao Zhang, Zhisheng Zheng, Wenxi Chen, Xiquan Li,
  Jiaxin Ye, Xie Chen, and Thomas Hain.
\newblock Emobox: Multilingual multi-corpus speech emotion recognition toolkit
  and benchmark.
\newblock In Itshak Lapidot and Sharon Gannot (eds.), \emph{25th Annual
  Conference of the International Speech Communication Association, Interspeech
  2024, Kos, Greece, September 1-5, 2024}. {ISCA}, 2024.
\newblock \doi{10.21437/INTERSPEECH.2024-788}.
\newblock URL \url{https://doi.org/10.21437/Interspeech.2024-788}.

\bibitem[Niizumi et~al.(2025)Niizumi, Takeuchi, Yasuda, Nguyen, Ohishi, and
  Harada]{niizumi2025m2d2}
Daisuke Niizumi, Daiki Takeuchi, Masahiro Yasuda, Binh~Thien Nguyen, Yasunori
  Ohishi, and Noboru Harada.
\newblock {M2D-CLAP:} exploring general-purpose audio-language representations
  beyond {CLAP}.
\newblock \emph{{IEEE} Access}, 13:\penalty0 163313--163330, 2025.
\newblock \doi{10.1109/ACCESS.2025.3611348}.
\newblock URL \url{https://doi.org/10.1109/ACCESS.2025.3611348}.

\bibitem[OpenAI(2024)]{hurst2024gpt}
OpenAI.
\newblock Gpt-4o system card.
\newblock \emph{CoRR}, abs/2410.21276, 2024.
\newblock \doi{10.48550/ARXIV.2410.21276}.
\newblock URL \url{https://doi.org/10.48550/arXiv.2410.21276}.

\bibitem[Osman et~al.(2024)Osman, Kaplan, and Nadeem]{osman_et_al_2024}
Mohamed Osman, Daniel~Z. Kaplan, and Tamer Nadeem.
\newblock {SER} evals: In-domain and out-of-domain benchmarking for speech
  emotion recognition.
\newblock In Itshak Lapidot and Sharon Gannot (eds.), \emph{25th Annual
  Conference of the International Speech Communication Association, Interspeech
  2024, Kos, Greece, September 1-5, 2024}. {ISCA}, 2024.
\newblock \doi{10.21437/INTERSPEECH.2024-2440}.
\newblock URL \url{https://doi.org/10.21437/Interspeech.2024-2440}.

\bibitem[Plutchik(2001)]{plutchik2001nature}
Robert Plutchik.
\newblock The nature of emotions: Human emotions have deep evolutionary roots,
  a fact that may explain their complexity and provide tools for clinical
  practice.
\newblock \emph{American scientist}, 89\penalty0 (4):\penalty0 344--350, 2001.

\bibitem[Radford et~al.(2023)Radford, Kim, Xu, Brockman, McLeavey, and
  Sutskever]{radford2023whisper}
Alec Radford, Jong~Wook Kim, Tao Xu, Greg Brockman, Christine McLeavey, and
  Ilya Sutskever.
\newblock Robust speech recognition via large-scale weak supervision.
\newblock In Andreas Krause, Emma Brunskill, Kyunghyun Cho, Barbara Engelhardt,
  Sivan Sabato, and Jonathan Scarlett (eds.), \emph{International Conference on
  Machine Learning, {ICML} 2023, 23-29 July 2023, Honolulu, Hawaii, {USA}},
  volume 202 of \emph{Proceedings of Machine Learning Research}, pp.\
  28492--28518. {PMLR}, 2023.
\newblock URL \url{https://proceedings.mlr.press/v202/radford23a.html}.

\bibitem[Reimers \& Gurevych(2019)Reimers and Gurevych]{reimers2019sentence}
Nils Reimers and Iryna Gurevych.
\newblock Sentence-bert: Sentence embeddings using siamese bert-networks.
\newblock In Kentaro Inui, Jing Jiang, Vincent Ng, and Xiaojun Wan (eds.),
  \emph{Proceedings of the 2019 Conference on Empirical Methods in Natural
  Language Processing and the 9th International Joint Conference on Natural
  Language Processing, {EMNLP-IJCNLP} 2019, Hong Kong, China, November 3-7,
  2019}, pp.\  3980--3990. Association for Computational Linguistics, 2019.
\newblock \doi{10.18653/V1/D19-1410}.
\newblock URL \url{https://doi.org/10.18653/v1/D19-1410}.

\bibitem[Rum \& Perry(2020)Rum and Perry]{rum2020empathic}
Yonat Rum and Anat Perry.
\newblock Empathic accuracy in clinical populations.
\newblock \emph{Frontiers in Psychiatry}, 11:\penalty0 457, 2020.

\bibitem[Russell(1980)]{russell1980circumplex}
James~A Russell.
\newblock A circumplex model of affect.
\newblock \emph{Journal of personality and social psychology}, 39\penalty0
  (6):\penalty0 1161, 1980.

\bibitem[Scheidwasser{-}Clow et~al.(2022)Scheidwasser{-}Clow, Kegler, Beckmann,
  and Cernak]{scheidwasser_clow_et_al_2021}
Neil Scheidwasser{-}Clow, Mikolaj Kegler, Pierre Beckmann, and Milos Cernak.
\newblock {SERAB:} {A} multi-lingual benchmark for speech emotion recognition.
\newblock In \emph{{IEEE} International Conference on Acoustics, Speech and
  Signal Processing, {ICASSP} 2022, Virtual and Singapore, 23-27 May 2022},
  pp.\  7697--7701. {IEEE}, 2022.
\newblock \doi{10.1109/ICASSP43922.2022.9747348}.
\newblock URL \url{https://doi.org/10.1109/ICASSP43922.2022.9747348}.

\bibitem[Schuhmann et~al.(2025{\natexlab{a}})Schuhmann, Kaczmarczyk, Rabby,
  Friedrich, Kraus, Nadi, Nguyen, Kersting, and
  Auer]{schuhmann2026emonetvoicefinegrainedexpertverifiedbenchmark}
Christoph Schuhmann, Robert Kaczmarczyk, Gollam Rabby, Felix Friedrich, Maurice
  Kraus, Kourosh Nadi, Huu Nguyen, Kristian Kersting, and S{\"{o}}ren Auer.
\newblock Emonet-voice: {A} fine-grained, expert-verified benchmark for speech
  emotion detection, 2025{\natexlab{a}}.
\newblock URL \url{https://doi.org/10.48550/arXiv.2506.09827}.

\bibitem[Schuhmann et~al.(2025{\natexlab{b}})Schuhmann, Kaczmarczyk, Rabby,
  Kraus, Friedrich, Nguyen, Kalyan, Nadi, Kersting, and
  Auer]{schuhmann2025emonetface}
Christoph Schuhmann, Robert Kaczmarczyk, Gollam Rabby, Maurice Kraus, Felix
  Friedrich, Huu Nguyen, Krishna Kalyan, Kourosh Nadi, Kristian Kersting, and
  S{\"{o}}ren Auer.
\newblock Emonet-face: An expert-annotated benchmark for synthetic emotion
  recognition.
\newblock In Danielle Belgrave, Cheng Zhang, Laura~N. Montoya, Hsuan{-}Tien
  Lin, Razvan Pascanu, Piotr Koniusz, Marzyeh Ghassemi, Nancy Chen, Iv{\'{a}}n
  Vladimir~Meza Ru{\'{\i}}z, and Arturo Loaiza{-}Bonilla (eds.), \emph{Advances
  in Neural Information Processing Systems 38: Annual Conference on Neural
  Information Processing Systems 2025, NeurIPS 2025, San Diego, CA, USA,
  December 2-7, 2025 / Mexico City, Mexico, November 30 - December 5, 2025},
  2025{\natexlab{b}}.
\newblock URL
  \url{http://papers.nips.cc/paper\_files/paper/2025/hash/8b2dd02001e72f65091115fb7f39d8c7-Abstract-Datasets\_and\_Benchmarks\_Track.html}.

\bibitem[Schuller(2018)]{schuller2018voice}
Bj{\"{o}}rn~W. Schuller.
\newblock Speech emotion recognition: two decades in a nutshell, benchmarks,
  and ongoing trends.
\newblock \emph{Commun. {ACM}}, 61\penalty0 (5):\penalty0 90--99, 2018.
\newblock \doi{10.1145/3129340}.
\newblock URL \url{https://doi.org/10.1145/3129340}.

\bibitem[Schuller et~al.(2013)Schuller, Steidl, Batliner, Vinciarelli, Scherer,
  Ringeval, Chetouani, Weninger, Eyben, Marchi, Mortillaro, Salamin,
  Polychroniou, Valente, and Kim]{schuller2013-kn}
Bj{\"{o}}rn~W. Schuller, Stefan Steidl, Anton Batliner, Alessandro Vinciarelli,
  Klaus~R. Scherer, Fabien Ringeval, Mohamed Chetouani, Felix Weninger, Florian
  Eyben, Erik Marchi, Marcello Mortillaro, Hugues Salamin, Anna Polychroniou,
  Fabio Valente, and Samuel Kim.
\newblock The {INTERSPEECH} 2013 computational paralinguistics challenge:
  social signals, conflict, emotion, autism.
\newblock In Fr{\'{e}}d{\'{e}}ric Bimbot, Christophe Cerisara, C{\'{e}}cile
  Fougeron, Guillaume Gravier, Lori Lamel, Fran{\c{c}}ois Pellegrino, and
  Pascal Perrier (eds.), \emph{14th Annual Conference of the International
  Speech Communication Association, {INTERSPEECH} 2013, Lyon, France, August
  25-29, 2013}, pp.\  148--152. {ISCA}, 2013.
\newblock \doi{10.21437/INTERSPEECH.2013-56}.
\newblock URL \url{https://doi.org/10.21437/Interspeech.2013-56}.

\bibitem[Stappen et~al.(2021)Stappen, Baird, Christ, Schumann, Sertolli,
  Me{\ss}ner, Cambria, Zhao, and Schuller]{stappen2021muse}
Lukas Stappen, Alice Baird, Lukas Christ, Lea Schumann, Benjamin Sertolli,
  Eva{-}Maria Me{\ss}ner, Erik Cambria, Guoying Zhao, and Bj{\"{o}}rn~W.
  Schuller.
\newblock The muse 2021 multimodal sentiment analysis challenge: Sentiment,
  emotion, physiological-emotion, and stress.
\newblock In Bj{\"{o}}rn~W. Schuller, Lukas Stappen, Eva{-}Maria Me{\ss}ner,
  Erik Cambria, and Guoying Zhao (eds.), \emph{MuSe '21: Proceedings of the 2nd
  on Multimodal Sentiment Analysis Challenge, Virtual Event, China, 24 October
  2021}, pp.\  5--14. {ACM}, 2021.
\newblock \doi{10.1145/3475957.3484450}.
\newblock URL \url{https://doi.org/10.1145/3475957.3484450}.

\bibitem[Tutt{\"{o}}s{\'{\i}} et~al.(2026)Tutt{\"{o}}s{\'{\i}}, Dhillon, Sang,
  Eastwood, Bhatia, Dinh, Kapoor, Jin, and Lim]{tuttosi_et_al_2025}
Paige Tutt{\"{o}}s{\'{\i}}, Mantaj Dhillon, Luna Sang, Shane Eastwood, Poorvi
  Bhatia, Quang~Minh Dinh, Avni Kapoor, Yewon Jin, and Angelica Lim.
\newblock Bersting at the screams: {A} benchmark for distanced, emotional and
  shouted speech recognition.
\newblock \emph{Comput. Speech Lang.}, 95:\penalty0 101815, 2026.
\newblock \doi{10.1016/J.CSL.2025.101815}.
\newblock URL \url{https://doi.org/10.1016/j.csl.2025.101815}.

\bibitem[van~den Oord et~al.(2018)van~den Oord, Li, and
  Vinyals]{oord2018representation}
A{\"{a}}ron van~den Oord, Yazhe Li, and Oriol Vinyals.
\newblock Representation learning with contrastive predictive coding.
\newblock \emph{CoRR}, abs/1807.03748, 2018.
\newblock URL \url{http://arxiv.org/abs/1807.03748}.

\bibitem[Wang et~al.(2020)Wang, Wei, Dong, Bao, Yang, and Zhou]{wang2020minilm}
Wenhui Wang, Furu Wei, Li~Dong, Hangbo Bao, Nan Yang, and Ming Zhou.
\newblock Minilm: Deep self-attention distillation for task-agnostic
  compression of pre-trained transformers.
\newblock 2020.
\newblock URL
  \url{https://proceedings.neurips.cc/paper/2020/hash/3f5ee243547dee91fbd053c1c4a845aa-Abstract.html}.

\bibitem[Wilting et~al.(2006)Wilting, Krahmer, and Swerts]{wilting2006real}
Janneke Wilting, Emiel Krahmer, and Marc Swerts.
\newblock Real vs. acted emotional speech.
\newblock In \emph{Ninth International Conference on Spoken Language
  Processing, {INTERSPEECH-ICSLP} 2006, Pittsburgh, PA, USA, September 17-21,
  2006}. {ISCA}, 2006.
\newblock \doi{10.21437/INTERSPEECH.2006-276}.
\newblock URL \url{https://doi.org/10.21437/Interspeech.2006-276}.

\bibitem[Wu et~al.(2023)Wu, Chen, Zhang, Hui, Berg{-}Kirkpatrick, and
  Dubnov]{wu2023clap}
Yusong Wu, Ke~Chen, Tianyu Zhang, Yuchen Hui, Taylor Berg{-}Kirkpatrick, and
  Shlomo Dubnov.
\newblock Large-scale contrastive language-audio pretraining with feature
  fusion and keyword-to-caption augmentation.
\newblock In \emph{{IEEE} International Conference on Acoustics, Speech and
  Signal Processing {ICASSP} 2023, Rhodes Island, Greece, June 4-10, 2023},
  pp.\  1--5. {IEEE}, 2023.
\newblock \doi{10.1109/ICASSP49357.2023.10095969}.
\newblock URL \url{https://doi.org/10.1109/ICASSP49357.2023.10095969}.

\bibitem[Xiao et~al.(2025)Xiao, Chan, Zhang, Xu, Aljunied, and
  Rong]{xiao2025scaling}
Chenghao Xiao, Hou~Pong Chan, Hao Zhang, Weiwen Xu, Mahani Aljunied, and
  Yu~Rong.
\newblock Scaling language-centric omnimodal representation learning.
\newblock In Danielle Belgrave, Cheng Zhang, Laura~N. Montoya, Hsuan{-}Tien
  Lin, Razvan Pascanu, Piotr Koniusz, Marzyeh Ghassemi, Nancy Chen, Iv{\'{a}}n
  Vladimir~Meza Ru{\'{\i}}z, and Arturo Loaiza{-}Bonilla (eds.), \emph{Advances
  in Neural Information Processing Systems 38: Annual Conference on Neural
  Information Processing Systems 2025, NeurIPS 2025, San Diego, CA, USA,
  December 2-7, 2025 / Mexico City, Mexico, November 30 - December 5, 2025},
  2025.
\newblock URL
  \url{http://papers.nips.cc/paper\_files/paper/2025/hash/e839f274b3d3d682a0a82e5129d2cc4a-Abstract-Conference.html}.

\bibitem[Yang et~al.(2026{\natexlab{a}})Yang, Yu, Chen, Zhu, Chen, Chen, Wang,
  Wang, Jiang, Jiang, Lin, Chen, Fei, Liu, Yu, Zhan, Yu, Huang, Fan, Chen,
  Cheng, Li, Li, Wang, Zhao, Gao, Gong, Zhang, Xu, and Qiu]{mossaudio2026}
Chen Yang, Chufan Yu, Hanfu Chen, Jie Zhu, Jingqi Chen, Ke~Chen, Wenxuan Wang,
  Yang Wang, Yaozhou Jiang, Yi~Jiang, Zhengyuan Lin, Ziqi Chen, Zhaoye Fei,
  Chenghao Liu, Donghua Yu, Jun Zhan, Kang Yu, Kexin Huang, Liwei Fan, Mingshu
  Chen, Qinyuan Cheng, Ruixiao Li, Shimin Li, Songlin Wang, Xingjian Zhao, Yang
  Gao, Yitian Gong, Yiyang Zhang, Zhe Xu, and Xipeng Qiu.
\newblock Moss-audio technical report, 2026{\natexlab{a}}.
\newblock URL \url{https://doi.org/10.48550/arXiv.2606.01802}.

\bibitem[Yang et~al.(2025)Yang, Yang, Jin, Cui, Wu, Li, Zhang, and
  Woodland]{yang2025spear}
Xiaoyu Yang, Yifan Yang, Zengrui Jin, Ziyun Cui, Wen Wu, Baoxiang Li, Chao
  Zhang, and Phil Woodland.
\newblock Spear: A unified ssl framework for learning speech and audio
  representations.
\newblock 2025.

\bibitem[Yang et~al.(2026{\natexlab{b}})Yang, Han, Wang, Wang, Ma, Zhou, Jin,
  Yang, Wang, Tan, and Chen]{yang2026towards}
Yifan Yang, Bing Han, Hui Wang, Wei Wang, Ziyang Ma, Long Zhou, Zengrui Jin,
  Guanrou Yang, Tianrui Wang, Xu~Tan, and Xie Chen.
\newblock Towards fine-grained and multi-granular contrastive language-speech
  pre-training.
\newblock In Maria Liakata, Viviane~P. Moreira, Jiajun Zhang, and David Jurgens
  (eds.), \emph{Proceedings of the 64th Annual Meeting of the Association for
  Computational Linguistics (Volume 1: Long Papers), {ACL} 2026, San Diego,
  California, United States, July 2-7, 2026}, pp.\  4217--4235. Association for
  Computational Linguistics, 2026{\natexlab{b}}.
\newblock \doi{10.18653/V1/2026.ACL-LONG.194}.
\newblock URL \url{https://doi.org/10.18653/v1/2026.acl-long.194}.

\bibitem[Zaki et~al.(2009)Zaki, Bolger, and Ochsner]{zaki2009empathic}
Jamil Zaki, Niall Bolger, and Kevin Ochsner.
\newblock Unpacking the informational bases of empathic accuracy.
\newblock \emph{Emotion}, 9\penalty0 (4):\penalty0 478, 2009.

\bibitem[Zhai et~al.(2023)Zhai, Mustafa, Kolesnikov, and
  Beyer]{zhai2023sigmoid}
Xiaohua Zhai, Basil Mustafa, Alexander Kolesnikov, and Lucas Beyer.
\newblock Sigmoid loss for language image pre-training.
\newblock In \emph{{IEEE/CVF} International Conference on Computer Vision,
  {ICCV} 2023, Paris, France, October 1-6, 2023}, pp.\  11941--11952. {IEEE},
  2023.
\newblock \doi{10.1109/ICCV51070.2023.01100}.
\newblock URL \url{https://doi.org/10.1109/ICCV51070.2023.01100}.

\bibitem[Zhang et~al.(2020)Zhang, Ju, Li, Li, Zhu, and Zhou]{zhang2020multi}
Dong Zhang, Xincheng Ju, Junhui Li, Shoushan Li, Qiaoming Zhu, and Guodong
  Zhou.
\newblock Multi-modal multi-label emotion detection with modality and label
  dependence.
\newblock In Bonnie Webber, Trevor Cohn, Yulan He, and Yang Liu (eds.),
  \emph{Proceedings of the 2020 Conference on Empirical Methods in Natural
  Language Processing, {EMNLP} 2020, Online, November 16-20, 2020}, pp.\
  3584--3593. Association for Computational Linguistics, 2020.
\newblock \doi{10.18653/V1/2020.EMNLP-MAIN.291}.
\newblock URL \url{https://doi.org/10.18653/v1/2020.emnlp-main.291}.

\bibitem[Zhu et~al.(2024)Zhu, Darefsky, and Duan]{zhu2024cacophony}
Ge~Zhu, Jordan Darefsky, and Zhiyao Duan.
\newblock Cacophony: An improved contrastive audio-text model.
\newblock \emph{{IEEE} {ACM} Trans. Audio Speech Lang. Process.}, 32:\penalty0
  4867--4879, 2024.
\newblock \doi{10.1109/TASLP.2024.3485170}.
\newblock URL \url{https://doi.org/10.1109/TASLP.2024.3485170}.

\end{thebibliography}

\clearpage
\appendix
\onecolumn

\section{Appendices}

\subsection{Comparison to Prior Speech-Emotion Benchmarks}
\label{app:ser_datasets_comparison}

Table~\ref{tab:ser_datasets_reformatted} situates \voicenetemo and \voicenetext against prior speech-emotion benchmarks. The table covers benchmarks only; the MOSS-Audio-annotated training corpora (\emoliabal, LAION's Got Talent, Majestrino, Multilingual In The Wild) are summarised separately in App.~\ref{app:emolia_corpora}.

\begin{table*}[h]
    \centering
    \scriptsize
    \caption{\textbf{Comparison of speech-emotion benchmarks}. \#Labels count the size of the label set: emotion or affect categories for prior benchmarks and \voicenetemo, talking-style attributes for \voicenetext. Real = recorded human speech; In-the-wild = unscripted, naturalistic. Open license means CC-BY or CC-BY-NC 4.0 or equivalent; var.~means varies across pooled corpora.}
    \label{tab:ser_datasets_reformatted}
    \setlength{\tabcolsep}{2pt}
    \resizebox{\linewidth}{!}{%
    \begin{tabular}{@{}p{0.8em}@{}lcccccc@{}}
    \toprule
    &\textbf{Dataset} & \shortstack{\textbf{Open}\\\textbf{Licence}} & \shortstack{\textbf{Size}\\\textbf{(\#Utts/Hours)}} & \textbf{\#Labels} & \textbf{\#Spk.} & \textbf{Real / In-wild} & \textbf{Multilin.}\\
    \midrule
    &IEMOCAP~\citep{busso2008iemocap} & \textcolor{red}{\ding{55}} & 10k / $\sim$12h & \phantom{$\le$}9 & 10 (5M/5F) & R / \textcolor{red}{\ding{55}} & \textcolor{red}{\ding{55}}\\
    &RAVDESS~\citep{livingstone2018ravdess} & \textcolor{green}{\ding{51}} & 1.4k / $\sim$1h & \phantom{$\le$}8 & 24 (12M/12F) & R / \textcolor{red}{\ding{55}} & \textcolor{red}{\ding{55}}\\
    &SAVEE~\citep{jackson2014savee} & \textcolor{red}{\ding{55}} & 480 / $<$1h & \phantom{$\le$}7 & 4 (Male) & R / \textcolor{red}{\ding{55}} & \textcolor{red}{\ding{55}}\\
    &EmoDB~\citep{burkhardt2005emodb} & \textcolor{red}{\ding{55}} & 535 / $<$1h & \phantom{$\le$}7 & 10 (5M/5F) & R / \textcolor{red}{\ding{55}} & \textcolor{red}{\ding{55}}\\
    &CREMA-D~\citep{cao2014crema} & \textcolor{green}{\ding{51}} & 7.4k / $\sim$6h & \phantom{$\le$}6 & 91 (48M/43F) & R / \textcolor{red}{\ding{55}} & \textcolor{red}{\ding{55}}\\
    &SERAB~\citep{scheidwasser_clow_et_al_2021} & \textcolor{red}{\ding{55}} & 9 corpora / var. & \phantom{$\le$}6 & var. & R / \textcolor{red}{\ding{55}} & \textcolor{green}{\ding{51}}\\
    &EmoBox~\citep{ma_et_al_2024} & \textcolor{red}{\ding{55}} & 32 corpora / var. & $\le$8 & var. & R / partial & \textcolor{green}{\ding{51}}\\
    &SER~Evals~\citep{osman_et_al_2024} & \textcolor{red}{\ding{55}} & 18 corpora / var. & $\le$8 & var. & R / partial & \textcolor{green}{\ding{51}}\\
    &{MSP-Podcast~\citep{lotfian2019msp}} & {\textcolor{green}{\ding{51}}} & {$\sim$100k / $\sim$100h} & 4 & {var.} & R / \textcolor{green}{\ding{51}} & {\textcolor{red}{\ding{55}}}\\
    &BERSt~\citep{tuttosi_et_al_2025} & \textcolor{green}{\ding{51}} & $\sim$4h & \phantom{$\le$}6 & 98 & R / \textcolor{red}{\ding{55}} & \textcolor{red}{\ding{55}}\\
    &\emonet~Bench \citep{schuhmann2026emonetvoicefinegrainedexpertverifiedbenchmark} & \textcolor{green}{\ding{51}} & $\sim$12k / $\sim$36h & \textbf{40} & 11 (Synth) & Synth / \textcolor{red}{\ding{55}} & \textcolor{green}{\ding{51}}\\
    \cmidrule{1-8}
    \multirow{3}{*}{\rotatebox{90}{ours}}
    &\textbf{\voicenetemo} (bench) & \textcolor{green}{\ding{51}} & \textbf{7{,}988 pairs / 3{,}944 clips} & \textbf{40} & var. & R / \textcolor{green}{\ding{51}} & \textcolor{green}{\ding{51}}\\
    &\textbf{\voicenetext} (bench) & \textcolor{green}{\ding{51}} & \textbf{18.5k pairs} & \textbf{57} & var. & R / \textcolor{green}{\ding{51}} & \textcolor{green}{\ding{51}}\\
    \bottomrule
    \end{tabular}
    }
\end{table*}

\subsection{\emolia and \emoliabal Corpora}
\label{app:emolia_corpora}

\emolia is the fully emotion-annotated version of the Emilia-Large corpus (71.78M clips, 215{,}600 hours; Emilia portion CC-BY-NC-4.0, Emilia-YODAS portion CC-BY-4.0), annotated with 40 emotion scores, emotion captions, and speaker-timbre embeddings. \emoliabal is a 5.26M-clip balanced subset derived from \emolia. Table~\ref{tab:emolia_corpora} lists the four corpora that received dense MOSS-Audio annotations.

\begin{table}[h]
\centering
\small
\caption{The four \textsc{MOSS-Audio}-annotated voice corpora. Each clip is queried with 18 prompt-groups; the per-corpus annotation-call counts follow.}
\label{tab:emolia_corpora}
\setlength{\tabcolsep}{6pt}
\begin{tabular}{@{}lrr@{}}
\toprule
Source dataset & Clips & Annotation calls (18\,$\times$) \\
\midrule
Multilingual In The Wild & 721k & 12.98M \\
Majestrino & 973k & 17.51M \\
LAION's Got Talent & 1.69M & 30.34M \\
\textbf{\emoliabal} (balanced subset of \emolia) & \textbf{5.26M} & \textbf{94.62M} \\
\midrule
\textbf{Total} & \textbf{8.64M} & \textbf{155.46M} \\
\bottomrule
\end{tabular}
\end{table}

\subsection{\emonet Taxonomy}
\label{app:taxonomy}

The 40 emotion categories used in \emonet, adapted from \emonetface \citep{schuhmann2025emonetface}, are listed below with associated descriptive terms used during conceptualization and prompting:

\begin{itemize}[noitemsep,topsep=0pt]
    \item \textbf{Amusement:} `lighthearted fun', `amusement', `mirth', `joviality', `laughter', `playfulness', `silliness', and `jesting'
    \item \textbf{Elation:} `happiness', `excitement', `joy', `exhilaration', `delight', `jubilation', `bliss', and `Cheerfulness'
    \item \textbf{Pleasure/Ecstasy:} `ecstasy', `pleasure', `bliss', `rapture', and `Beatitude'
    \item \textbf{Contentment:} `contentment', `relaxation', `peacefulness', `calmness', `satisfaction', `Ease', `Serenity', `fulfillment', `gladness', `lightness', `serenity', and `tranquility'
    \item \textbf{Thankfulness/Gratitude:} `thankfulness', `gratitude', `appreciation', and `gratefulness'
    \item \textbf{Affection:} `sympathy', `compassion', `warmth', `trust', `caring', `Clemency', `forgiveness', `Devotion', `Tenderness', and `Reverence'
    \item \textbf{Infatuation:} `infatuation', `having a crush', `romantic desire', `fondness', `butterflies in the stomach', and `adoration'
    \item \textbf{Hope/Enthusiasm/Optimism:} `hope', `enthusiasm', `optimism', `Anticipation', `Courage', `Encouragement', `Zeal', `fervor', `inspiration', and `Determination'
    \item \textbf{Triumph:} `triumph', `superiority'
    \item \textbf{Pride:} `pride', `dignity', `self-confidently', `honor', and `self-consciousness'
    \item \textbf{Interest:} `interest', `fascination', `curiosity', and `intrigue'
    \item \textbf{Awe:} `awe', `awestruck', and `wonder'
    \item \textbf{Astonishment/Surprise:} `astonishment', `surprise', `amazement', `shock', and `startlement'
    \item \textbf{Concentration:} `concentration', `deep focus', `engrossment', `absorption', and `attention'
    \item \textbf{Contemplation:} `contemplation', `thoughtfulness', `pondering', `reflection', `meditation', `Brooding', and `Pensiveness'
    \item \textbf{Relief:} `relief', `respite', `alleviation', `solace', `comfort', and `liberation'
    \item \textbf{Longing:} `yearning', `longing', `pining', `wistfulness', `nostalgia', `Craving', `desire', `Envy', `homesickness', and `saudade'
    \item \textbf{Teasing:} `teasing', `bantering', `mocking playfully', `ribbing', and `provoking lightly'
    \item \textbf{Impatience and Irritability:} `impatience', `irritability', `irritation', `restlessness', `short-temperedness', and `exasperation'
    \item \textbf{Sexual Lust:} `sexual lust', `carnal desire', `lust', `feeling horny', and `feeling turned on'
    \item \textbf{Doubt:} `doubt', `distrust', `suspicion', `skepticism', `uncertainty', and `Pessimism'
    \item \textbf{Fear:} `fear', `terror', `dread', `apprehension', `alarm', `horror', `panic', and `nervousness'
    \item \textbf{Distress:} `worry', `anxiety', `unease', `anguish', `trepidation', `Concern', `Upset', `pessimism', and `foreboding'
    \item \textbf{Confusion:} `confusion', `bewilderment', `flabbergasted', `disorientation', and `Perplexity'
    \item \textbf{Embarrassment:} `embarrassment', `shyness', `mortification', `discomfiture', `awkwardness', and `Self-Consciousness'
    \item \textbf{Shame:} `shame', `guilt', `remorse', `humiliation', and `contrition'
    \item \textbf{Disappointment:} `disappointment', `regret', `dismay', `letdown', and `chagrin'
    \item \textbf{Sadness:} `sadness', `sorrow', `grief', `melancholy', `Dejection', `Despair', `Self-Pity', `Sullenness', `heartache', `mournfulness', and `misery'
    \item \textbf{Bitterness:} `resentment', `acrimony', `bitterness', `cynicism', and `rancor'
    \item \textbf{Contempt:} `contempt', `disapproval', `scorn', `disdain', `loathing', and `Detestation'
    \item \textbf{Disgust:} `disgust', `revulsion', `repulsion', `abhorrence', and `loathing'
    \item \textbf{Anger:} `anger', `rage', `fury', `hate', `irascibility', `enragement', `Vexation', `Wrath', `Peevishness', and `Annoyance'
    \item \textbf{Malevolence/Malice:} `spite', `sadism', `malevolence', `malice', `desire to harm', and `schadenfreude'
    \item \textbf{Sourness:} `sourness', `tartness', `acidity', `acerbity', and `sharpness' (Note: Primarily gustatory, vocal correlates might be subtle reactions)
    \item \textbf{Pain:} `physical pain', `suffering', `torment', `ache', and `agony'
    \item \textbf{Helplessness:} `helplessness', `powerlessness', `desperation', and `submission'
    \item \textbf{Fatigue/Exhaustion:} `fatigue', `exhaustion', `weariness', `lethargy', `burnout', and `Weariness'
    \item \textbf{Emotional Numbness:} `numbness', `detachment', `insensitivity', `emotional blunting', `apathy', `existential void', `boredom', `stoicism', and `indifference'
    \item \textbf{Intoxication/Altered States of Consciousness:} `being drunk', `stupor', `intoxication', `disorientation', and `altered perception'
    \item \textbf{Jealousy \& Envy:} `jealousy', `envy', and `covetousness'
\end{itemize}

\subsection{\voicenetext Attribute List}
\label{app:dim_attributes}

The 57 talking-style attributes annotated in \voicenetext are listed below in groups. Each attribute is shown with its MOSS-Audio short code in \texttt{typewriter} (4-letter, with an \texttt{R\_} prefix for resonance placement and an \texttt{S\_} prefix for style descriptors) followed by a plain-text description in parentheses; this is the schema used internally by the annotation pipeline (\S\ref{sec:methods_emolia}).

\textbf{Perceived speaker.} \texttt{GEND} (gender presentation) and \texttt{AGEV} (age range).

\textbf{Affective dimensions.} \texttt{VALN} (valence), \texttt{AROU} (arousal), \texttt{VOLT} (volatility), \texttt{VALS} (valence sharpness), \texttt{TENS} (vocal tension), \texttt{ARSH} (acoustic harshness), and \texttt{VULN} (vulnerability).

\textbf{Prosodic delivery.} \texttt{TEMP} (speaking rate), \texttt{ATCK} (attack), \texttt{CHNK} (chunking), \texttt{RANG} (melodic range), \texttt{VFLX} (vocal flexibility), \texttt{STNC} (vocal stance), \texttt{EMPH} (emphasis), \texttt{DFLU} (disfluency), \texttt{CLRT} (clarity), \texttt{STRU} (discourse structure), \texttt{COGL} (cognitive load), and \texttt{FOCS} (focus).

\textbf{Vocal quality.} \texttt{ROUG} (roughness), \texttt{SMTH} (smoothness), \texttt{BRGT} (brightness), \texttt{WARM} (warmth), \texttt{FULL} (fullness), \texttt{HARM} (harmonicity), \texttt{METL} (metalicity), \texttt{ESTH} (esthetics), \texttt{REGS} (register), \texttt{RESP} (respiration), \texttt{DARC} (dark/light), and \texttt{RCQL} (recording quality).

\textbf{Resonance placement.} \texttt{R\_CHST} (chest), \texttt{R\_THRT} (throat), \texttt{R\_ORAL} (oral), \texttt{R\_HEAD} (head), \texttt{R\_MASK} (mask), \texttt{R\_MIXD} (mixed), and \texttt{R\_NASL} (nasal).

\textbf{Recording context.} \texttt{BKGN} (background noise), \texttt{EXPL} (expletive content).

\textbf{Style descriptors.} \texttt{S\_CONV} (conversational), \texttt{S\_CASU} (casual), \texttt{S\_PLAY} (playful), \texttt{S\_CART} (cartoonish), \texttt{S\_FORM} (formal), \texttt{S\_AUTH} (authoritative), \texttt{S\_TECH} (teacher/didactic), \texttt{S\_MONO} (monologue), \texttt{S\_DRAM} (dramatic), \texttt{S\_NARR} (narrator), \texttt{S\_STRY} (storytelling), \texttt{S\_NEWS} (newsreader), \texttt{S\_RANT} (ranting), \texttt{S\_WHIS} (whisper-talk), and \texttt{S\_ASMR} (ASMR).

The MOSS-Audio schema additionally produces \texttt{BURST} (vocal-burst presence), \texttt{EMO} (fine-grained emotion category), \texttt{ACNT} (accent), and \texttt{LANG} (language), all dropped from \voicenetext because either redundant with \voicenetemo, demographic rather than perceptual, or categorical rather than ordinal. The full rubric-level definitions used for human annotation are available at \url{https://projects.laion.ai/emolia-bench/taxonomy/}.

\subsection{Dataset configuration}
\label{app:detailed_baseline_training}
Both \voiceclapsmall and \voiceclaplarge are trained on the same 9-corpus mixture. Each line of Tab.~\ref{tab:voiceclap_data_manifest} is one WebDataset entry with clip count and caption-key. \emoliabal, LAION's Got Talent, and Majestrino supply vocal-style captions; the four FCaps corpora (EARS, Expresso, Voxceleb1, and Voxceleb2) retain their prior FCaps-style captions \citep{yang2026towards}; the two Synthetic Vocal Bursts collections are sourced in-house and carry the same simpler caption schema. Multilingual In The Wild is annotated and released as part of the \emolia suite but excluded from the training mixture; an internal ablation showed it did not improve downstream \voicenetemo or MAEB-voice scores.

\begin{table}[h!]
\centering
\caption{\voiceclapdata training corpus manifest (Multilingual In The Wild is annotated and released as part of the \emolia suite but excluded from training after ablation). Caption-key indicates which JSON field is used as the contrastive text.}
\label{tab:voiceclap_data_manifest}
\setlength{\tabcolsep}{4pt}
\begin{tabular}{@{}lrl@{}}
\toprule
Corpus & Clips & Caption key \\
\midrule
LAION's Got Talent & 1{,}685{,}809 & \texttt{detailed\_caption} \\
\emoliabal & 5{,}256{,}683 & \texttt{emotion\_caption} \\
Majestrino & 972{,}658 & \texttt{caption} \\
Synthetic Vocal Bursts & 325{,}548 & \texttt{text} (in-house, YouTube) \\
Improved Synthetic Vocal Bursts & 15{,}680 & \texttt{text} (in-house, YouTube) \\
EARS & 17{,}227 & \texttt{text} (FCaps) \\
Expresso & 27{,}056 & \texttt{text} (FCaps) \\
Voxceleb1 & 153{,}516 & \texttt{text} (FCaps) \\
Voxceleb2 & 600{,}064 & \texttt{text} (FCaps) \\
\bottomrule
\end{tabular}
\vspace{0.2cm}
\end{table}

\subsection{Evaluation Harness and Loader Audit}
\label{app:loader_audit}

The harness implementation lives in \texttt{open\_clap\_scaling/\allowbreak eval\_emolia\_current\_\allowbreak\{clap,ensemble,baselines\}.py}. The function \texttt{evaluate\_subset} (\texttt{eval\_emolia\_current\_ensemble.py}) computes the similarity matrix, sweeps thresholds, and emits per-tag files alongside per-prompt and per-task-type CSVs. Per-prompt thresholds (\texttt{find\_per\_prompt\_thresholds}) require at least 10 rows per prompt; otherwise the global threshold is used.

\subsection{Per-tag Summary Files}
\label{app:full_summaries}

Full per-tag files (with \textit{bal@0}, \textit{bal@opt}, \textit{bal@per\_prompt}, mean per-prompt $\rho$, similarity-distribution stats, and per-prompt thresholds) are released alongside the paper. Numbers reported in this paper match those files exactly. Per-tag files additionally contain the per-(audio, prompt) cosine similarities and binary labels used for Tab.~\ref{tab:bootstrap_ci}. We draw 2,000 paired bootstrap samples with audio clip as the cluster, using the same sampled clips for every model. The 40 emotion and 395 retained attribute-level prompts are fixed; $\rho$ is recomputed within each prompt and then averaged. For \textit{bal@pp}, the Tab.~\ref{tab:model_results} thresholds are locked, so these intervals isolate clip-sampling uncertainty rather than refitting an oracle threshold in every replicate. The released script records the seed and input checksums and also emits paired intervals for every point-sorted adjacent model difference.

\begin{table}[h!]
\centering
\tiny
\setlength{\tabcolsep}{2.8pt}
\renewcommand{\arraystretch}{1.08}
\caption{Paired clip-bootstrap 95\% CIs for Tab.~\ref{tab:model_results} (2,000 replicates). Entries are point estimate $[\mathrm{lo},\mathrm{hi}]$. Per-prompt thresholds are held fixed; intervals therefore do not correct their same-set fitting. \voicenetext remains preliminary because this resampling also does not propagate annotation uncertainty.}
\label{tab:bootstrap_ci}
\resizebox{\textwidth}{!}{%
\begin{tabular}{@{}lcccc@{}}
\toprule
Model & \voicenetemo \textit{bal@pp} & \voicenetext \textit{bal@pp} & \voicenetemo $\rho$ & \voicenetext $\rho$ \\
\midrule
LAION-CLAP & 0.5624 [0.5511, 0.5727] & 0.6184 [0.6104, 0.6268] & 0.0310 [0.0071, 0.0537] & 0.0600 [0.0421, 0.0753] \\
MGA-CLAP & 0.5897 [0.5789, 0.5998] & 0.6208 [0.6128, 0.6287] & 0.0947 [0.0717, 0.1186] & 0.0819 [0.0653, 0.0962] \\
GLAP & 0.5765 [0.5661, 0.5877] & 0.6254 [0.6171, 0.6341] & 0.0667 [0.0425, 0.0907] & 0.0836 [0.0655, 0.0992] \\
M2D-CLAP-2025 & 0.5681 [0.5571, 0.5787] & 0.6361 [0.6273, 0.6444] & 0.0348 [0.0121, 0.0582] & 0.1003 [0.0822, 0.1142] \\
CLSP & 0.5675 [0.5563, 0.5791] & 0.6221 [0.6140, 0.6303] & 0.0610 [0.0380, 0.0840] & 0.0937 [0.0763, 0.1072] \\
MS-CLAP-2023 & 0.5754 [0.5646, 0.5858] & 0.6321 [0.6243, 0.6403] & $-$0.0084 [$-$0.0303, 0.0153] & 0.1044 [0.0867, 0.1186] \\
Cacophony & 0.5432 [0.5329, 0.5535] & 0.6089 [0.6005, 0.6173] & $-$0.0183 [$-$0.0414, 0.0034] & 0.0188 [0.0021, 0.0350] \\
LCO-Embedding-Omni-7B & 0.6632 [0.6525, 0.6740] & 0.6342 [0.6259, 0.6419] & 0.3052 [0.2824, 0.3266] & 0.0977 [0.0800, 0.1121] \\
LCO-Embedding-Omni-3B & 0.6547 [0.6435, 0.6656] & 0.6293 [0.6210, 0.6375] & 0.2876 [0.2646, 0.3082] & 0.0873 [0.0703, 0.1018] \\
\voiceclapsmall & 0.6754 [0.6648, 0.6860] & 0.6367 [0.6289, 0.6453] & 0.3176 [0.2932, 0.3391] & 0.1051 [0.0867, 0.1189] \\
\voiceclaplarge & \textbf{0.7021 [0.6914, 0.7118]} & \textbf{0.6510 [0.6431, 0.6588]} & \textbf{0.3719 [0.3487, 0.3923]} & \textbf{0.1475 [0.1292, 0.1599]} \\
\bottomrule
\end{tabular}}
\end{table}

\subsection{\voicenetemo Per-Emotion \texorpdfstring{$\rho$}{rho}}
\label{app:per_emotion}

\begin{table}[ht!]
    \centering
    \tiny
    \setlength{\tabcolsep}{2pt}
    \renewcommand{\arraystretch}{1.05}
    \caption{Per-emotion Spearman $\rho$ on \voicenetemo for \voiceclap, the Omni-Embedding bases, and all seven general-audio CLAP baselines. Rows are sorted by the mean $\rho$ across the displayed model columns; the mean is used only for ordering and is not shown. Best per row is bold, runner-up underlined. Cells are coloured by $\rho$.}
    \label{tab:model_results_by_emotion}
    \resizebox{\textwidth}{!}{%
\begin{tabular}{@{}p{3.6cm}*{11}{R{0.82cm}}@{}}
\toprule
\quad emotion & \multicolumn{1}{c}{\rotatebox[origin=lB]{90}{\makecell[l]{\textsc{VoiceCLAP}\\Small}}} & \multicolumn{1}{c}{\rotatebox[origin=lB]{90}{\makecell[l]{\textsc{VoiceCLAP}\\Large}}} & \multicolumn{1}{c}{\rotatebox[origin=lB]{90}{\makecell[l]{LCO\\Omni-3B}}} & \multicolumn{1}{c}{\rotatebox[origin=lB]{90}{\makecell[l]{LCO\\Omni-7B}}} & \multicolumn{1}{c}{\rotatebox[origin=lB]{90}{\makecell[l]{LAION\\CLAP}}} & \multicolumn{1}{c}{\rotatebox[origin=lB]{90}{\makecell[l]{MS-CLAP\\2023}}} & \multicolumn{1}{c}{\rotatebox[origin=lB]{90}{CLSP}} & \multicolumn{1}{c}{\rotatebox[origin=lB]{90}{GLAP}} & \multicolumn{1}{c}{\rotatebox[origin=lB]{90}{\makecell[l]{M2D\\CLAP}}} & \multicolumn{1}{c}{\rotatebox[origin=lB]{90}{\makecell[l]{MGA\\CLAP}}} & \multicolumn{1}{c}{\rotatebox[origin=lB]{90}{Cacophony}}\\
\midrule
Anger & \cellcolor[rgb]{0.667,0.400,0.733}\underline{0.555} & \cellcolor[rgb]{0.654,0.400,0.746}\textbf{0.577} & \cellcolor[rgb]{0.727,0.400,0.673}0.455 & \cellcolor[rgb]{0.686,0.400,0.714}0.524 & \cellcolor[rgb]{0.814,0.400,0.586}0.310 & \cellcolor[rgb]{0.791,0.400,0.609}0.348 & \cellcolor[rgb]{0.708,0.400,0.692}0.486 & \cellcolor[rgb]{0.767,0.400,0.633}0.389 & \cellcolor[rgb]{0.863,0.400,0.537}0.228 & \cellcolor[rgb]{0.731,0.400,0.669}0.448 & \cellcolor[rgb]{0.968,0.400,0.432}0.054 \\
Impatience and Irritability & \cellcolor[rgb]{0.617,0.400,0.783}\textbf{0.638} & \cellcolor[rgb]{0.651,0.400,0.749}\underline{0.581} & \cellcolor[rgb]{0.686,0.400,0.714}0.524 & \cellcolor[rgb]{0.674,0.400,0.726}0.544 & \cellcolor[rgb]{0.873,0.400,0.527}0.212 & \cellcolor[rgb]{1.000,0.400,0.400}$-$0.086 & \cellcolor[rgb]{0.911,0.400,0.489}0.149 & \cellcolor[rgb]{0.775,0.400,0.625}0.375 & \cellcolor[rgb]{1.000,0.400,0.400}$-$0.132 & \cellcolor[rgb]{0.707,0.400,0.693}0.489 & \cellcolor[rgb]{0.945,0.400,0.455}0.092 \\
Contempt & \cellcolor[rgb]{0.761,0.400,0.639}0.398 & \cellcolor[rgb]{0.735,0.400,0.665}\underline{0.441} & \cellcolor[rgb]{0.740,0.400,0.660}0.433 & \cellcolor[rgb]{0.702,0.400,0.698}\textbf{0.497} & \cellcolor[rgb]{0.862,0.400,0.538}0.230 & \cellcolor[rgb]{1.000,0.400,0.400}$-$0.008 & \cellcolor[rgb]{0.911,0.400,0.489}0.149 & \cellcolor[rgb]{0.848,0.400,0.552}0.254 & \cellcolor[rgb]{0.888,0.400,0.512}0.186 & \cellcolor[rgb]{0.764,0.400,0.636}0.394 & \cellcolor[rgb]{0.996,0.400,0.404}0.007 \\
Amusement & \cellcolor[rgb]{0.689,0.400,0.711}0.518 & \cellcolor[rgb]{0.700,0.400,0.700}0.500 & \cellcolor[rgb]{0.695,0.400,0.705}0.509 & \cellcolor[rgb]{0.682,0.400,0.718}\textbf{0.530} & \cellcolor[rgb]{0.830,0.400,0.570}0.283 & \cellcolor[rgb]{1.000,0.400,0.400}$-$0.281 & \cellcolor[rgb]{1.000,0.400,0.400}$-$0.351 & \cellcolor[rgb]{0.682,0.400,0.718}\underline{0.530} & \cellcolor[rgb]{0.963,0.400,0.437}0.062 & \cellcolor[rgb]{0.684,0.400,0.716}0.526 & \cellcolor[rgb]{0.908,0.400,0.492}0.154 \\
Distress & \cellcolor[rgb]{0.624,0.400,0.776}\textbf{0.627} & \cellcolor[rgb]{0.624,0.400,0.776}\underline{0.626} & \cellcolor[rgb]{0.677,0.400,0.723}0.538 & \cellcolor[rgb]{0.655,0.400,0.745}0.575 & \cellcolor[rgb]{1.000,0.400,0.400}$-$0.030 & \cellcolor[rgb]{1.000,0.400,0.400}$-$0.255 & \cellcolor[rgb]{0.864,0.400,0.536}0.226 & \cellcolor[rgb]{0.864,0.400,0.536}0.226 & \cellcolor[rgb]{0.805,0.400,0.595}0.325 & \cellcolor[rgb]{0.771,0.400,0.629}0.382 & \cellcolor[rgb]{1.000,0.400,0.400}$-$0.318 \\
Contemplation & \cellcolor[rgb]{0.731,0.400,0.669}\underline{0.449} & \cellcolor[rgb]{0.722,0.400,0.678}\textbf{0.464} & \cellcolor[rgb]{0.738,0.400,0.662}0.437 & \cellcolor[rgb]{0.746,0.400,0.654}0.423 & \cellcolor[rgb]{0.987,0.400,0.413}0.022 & \cellcolor[rgb]{0.813,0.400,0.587}0.312 & \cellcolor[rgb]{0.903,0.400,0.497}0.161 & \cellcolor[rgb]{1.000,0.400,0.400}$-$0.022 & \cellcolor[rgb]{0.917,0.400,0.483}0.138 & \cellcolor[rgb]{0.903,0.400,0.497}0.161 & \cellcolor[rgb]{0.889,0.400,0.511}0.185 \\
Disgust & \cellcolor[rgb]{0.812,0.400,0.588}0.313 & \cellcolor[rgb]{0.692,0.400,0.708}\underline{0.513} & \cellcolor[rgb]{0.702,0.400,0.698}0.497 & \cellcolor[rgb]{0.683,0.400,0.717}\textbf{0.528} & \cellcolor[rgb]{1.000,0.400,0.400}$-$0.031 & \cellcolor[rgb]{0.919,0.400,0.481}0.135 & \cellcolor[rgb]{0.897,0.400,0.503}0.172 & \cellcolor[rgb]{0.954,0.400,0.446}0.077 & \cellcolor[rgb]{0.901,0.400,0.499}0.165 & \cellcolor[rgb]{0.833,0.400,0.567}0.278 & \cellcolor[rgb]{1.000,0.400,0.400}$-$0.027 \\
Helplessness & \cellcolor[rgb]{0.734,0.400,0.666}\underline{0.444} & \cellcolor[rgb]{0.695,0.400,0.705}\textbf{0.508} & \cellcolor[rgb]{0.771,0.400,0.629}0.382 & \cellcolor[rgb]{0.749,0.400,0.651}0.418 & \cellcolor[rgb]{1.000,0.400,0.400}$-$0.262 & \cellcolor[rgb]{1.000,0.400,0.400}$-$0.015 & \cellcolor[rgb]{0.846,0.400,0.554}0.257 & \cellcolor[rgb]{0.850,0.400,0.550}0.250 & \cellcolor[rgb]{0.856,0.400,0.544}0.240 & \cellcolor[rgb]{0.789,0.400,0.611}0.351 & \cellcolor[rgb]{1.000,0.400,0.400}$-$0.214 \\
Malevolence/Malice & \cellcolor[rgb]{0.693,0.400,0.707}\textbf{0.512} & \cellcolor[rgb]{0.696,0.400,0.704}\underline{0.506} & \cellcolor[rgb]{0.746,0.400,0.654}0.424 & \cellcolor[rgb]{0.711,0.400,0.689}0.482 & \cellcolor[rgb]{0.908,0.400,0.492}0.153 & \cellcolor[rgb]{1.000,0.400,0.400}$-$0.015 & \cellcolor[rgb]{1.000,0.400,0.400}$-$0.109 & \cellcolor[rgb]{0.893,0.400,0.507}0.179 & \cellcolor[rgb]{0.955,0.400,0.445}0.075 & \cellcolor[rgb]{0.962,0.400,0.438}0.063 & \cellcolor[rgb]{1.000,0.400,0.400}$-$0.146 \\
Sadness & \cellcolor[rgb]{0.680,0.400,0.720}\underline{0.533} & \cellcolor[rgb]{0.659,0.400,0.741}\textbf{0.569} & \cellcolor[rgb]{0.771,0.400,0.629}0.381 & \cellcolor[rgb]{0.734,0.400,0.666}0.443 & \cellcolor[rgb]{1.000,0.400,0.400}$-$0.083 & \cellcolor[rgb]{0.873,0.400,0.527}0.212 & \cellcolor[rgb]{0.910,0.400,0.490}0.150 & \cellcolor[rgb]{1.000,0.400,0.400}$-$0.167 & \cellcolor[rgb]{0.876,0.400,0.524}0.206 & \cellcolor[rgb]{1.000,0.400,0.400}$-$0.042 & \cellcolor[rgb]{1.000,0.400,0.400}$-$0.194 \\
Confusion & \cellcolor[rgb]{0.781,0.400,0.619}0.365 & \cellcolor[rgb]{0.743,0.400,0.657}\textbf{0.429} & \cellcolor[rgb]{0.768,0.400,0.632}0.386 & \cellcolor[rgb]{0.765,0.400,0.635}\underline{0.391} & \cellcolor[rgb]{1.000,0.400,0.400}$-$0.063 & \cellcolor[rgb]{1.000,0.400,0.400}$-$0.190 & \cellcolor[rgb]{1.000,0.400,0.400}$-$0.017 & \cellcolor[rgb]{0.834,0.400,0.566}0.276 & \cellcolor[rgb]{1.000,0.400,0.400}$-$0.004 & \cellcolor[rgb]{0.871,0.400,0.529}0.215 & \cellcolor[rgb]{0.937,0.400,0.463}0.105 \\
Teasing & \cellcolor[rgb]{0.921,0.400,0.479}0.132 & \cellcolor[rgb]{0.778,0.400,0.622}0.370 & \cellcolor[rgb]{0.752,0.400,0.648}\underline{0.413} & \cellcolor[rgb]{0.699,0.400,0.701}\textbf{0.501} & \cellcolor[rgb]{0.841,0.400,0.559}0.265 & \cellcolor[rgb]{1.000,0.400,0.400}$-$0.276 & \cellcolor[rgb]{1.000,0.400,0.400}$-$0.150 & \cellcolor[rgb]{0.981,0.400,0.419}0.031 & \cellcolor[rgb]{1.000,0.400,0.400}$-$0.049 & \cellcolor[rgb]{0.779,0.400,0.621}0.369 & \cellcolor[rgb]{0.863,0.400,0.537}0.228 \\
Bitterness & \cellcolor[rgb]{0.783,0.400,0.617}\textbf{0.361} & \cellcolor[rgb]{0.812,0.400,0.588}\underline{0.314} & \cellcolor[rgb]{0.905,0.400,0.495}0.158 & \cellcolor[rgb]{0.847,0.400,0.553}0.255 & \cellcolor[rgb]{0.860,0.400,0.540}0.234 & \cellcolor[rgb]{0.900,0.400,0.500}0.166 & \cellcolor[rgb]{0.897,0.400,0.503}0.171 & \cellcolor[rgb]{1.000,0.400,0.400}$-$0.002 & \cellcolor[rgb]{0.899,0.400,0.501}0.169 & \cellcolor[rgb]{0.895,0.400,0.505}0.175 & \cellcolor[rgb]{1.000,0.400,0.400}$-$0.198 \\
Embarrassment & \cellcolor[rgb]{0.808,0.400,0.592}\underline{0.320} & \cellcolor[rgb]{0.805,0.400,0.595}\textbf{0.325} & \cellcolor[rgb]{0.841,0.400,0.559}0.265 & \cellcolor[rgb]{0.856,0.400,0.544}0.240 & \cellcolor[rgb]{0.968,0.400,0.432}0.053 & \cellcolor[rgb]{1.000,0.400,0.400}$-$0.049 & \cellcolor[rgb]{0.893,0.400,0.507}0.178 & \cellcolor[rgb]{0.888,0.400,0.512}0.187 & \cellcolor[rgb]{1.000,0.400,0.400}$-$0.002 & \cellcolor[rgb]{0.885,0.400,0.515}0.192 & \cellcolor[rgb]{0.975,0.400,0.425}0.042 \\
Triumph & \cellcolor[rgb]{0.825,0.400,0.575}\underline{0.292} & \cellcolor[rgb]{0.782,0.400,0.618}\textbf{0.363} & \cellcolor[rgb]{0.825,0.400,0.575}0.291 & \cellcolor[rgb]{0.837,0.400,0.563}0.271 & \cellcolor[rgb]{0.911,0.400,0.489}0.149 & \cellcolor[rgb]{0.961,0.400,0.439}0.065 & \cellcolor[rgb]{1.000,0.400,0.400}$-$0.108 & \cellcolor[rgb]{0.917,0.400,0.483}0.138 & \cellcolor[rgb]{1.000,0.400,0.400}$-$0.030 & \cellcolor[rgb]{0.877,0.400,0.523}0.205 & \cellcolor[rgb]{0.948,0.400,0.452}0.087 \\
Relief & \cellcolor[rgb]{0.833,0.400,0.567}0.279 & \cellcolor[rgb]{0.751,0.400,0.649}\textbf{0.415} & \cellcolor[rgb]{0.833,0.400,0.567}0.279 & \cellcolor[rgb]{0.803,0.400,0.597}\underline{0.328} & \cellcolor[rgb]{1.000,0.400,0.400}$-$0.122 & \cellcolor[rgb]{0.950,0.400,0.450}0.083 & \cellcolor[rgb]{0.849,0.400,0.551}0.251 & \cellcolor[rgb]{1.000,0.400,0.400}$-$0.056 & \cellcolor[rgb]{0.920,0.400,0.480}0.133 & \cellcolor[rgb]{0.972,0.400,0.428}0.046 & \cellcolor[rgb]{0.972,0.400,0.428}0.046 \\
Disappointment & \cellcolor[rgb]{0.807,0.400,0.593}0.321 & \cellcolor[rgb]{0.774,0.400,0.626}\underline{0.377} & \cellcolor[rgb]{0.768,0.400,0.632}\textbf{0.386} & \cellcolor[rgb]{0.779,0.400,0.621}0.368 & \cellcolor[rgb]{1.000,0.400,0.400}$-$0.071 & \cellcolor[rgb]{0.892,0.400,0.508}0.180 & \cellcolor[rgb]{0.930,0.400,0.470}0.116 & \cellcolor[rgb]{0.965,0.400,0.435}0.058 & \cellcolor[rgb]{0.900,0.400,0.500}0.166 & \cellcolor[rgb]{1.000,0.400,0.400}$-$0.147 & \cellcolor[rgb]{1.000,0.400,0.400}$-$0.093 \\
Fatigue/Exhaustion & \cellcolor[rgb]{0.707,0.400,0.693}\underline{0.489} & \cellcolor[rgb]{0.697,0.400,0.703}\textbf{0.505} & \cellcolor[rgb]{0.783,0.400,0.617}0.362 & \cellcolor[rgb]{0.792,0.400,0.608}0.347 & \cellcolor[rgb]{1.000,0.400,0.400}$-$0.135 & \cellcolor[rgb]{1.000,0.400,0.400}$-$0.054 & \cellcolor[rgb]{1.000,0.400,0.400}$-$0.028 & \cellcolor[rgb]{0.951,0.400,0.449}0.082 & \cellcolor[rgb]{0.907,0.400,0.493}0.155 & \cellcolor[rgb]{0.995,0.400,0.405}0.009 & \cellcolor[rgb]{1.000,0.400,0.400}$-$0.083 \\
Thankfulness/Gratitude & \cellcolor[rgb]{0.959,0.400,0.441}0.068 & \cellcolor[rgb]{0.849,0.400,0.551}0.252 & \cellcolor[rgb]{0.789,0.400,0.611}\textbf{0.351} & \cellcolor[rgb]{0.816,0.400,0.584}\underline{0.307} & \cellcolor[rgb]{1.000,0.400,0.400}$-$0.003 & \cellcolor[rgb]{0.988,0.400,0.412}0.020 & \cellcolor[rgb]{0.919,0.400,0.481}0.135 & \cellcolor[rgb]{0.918,0.400,0.482}0.136 & \cellcolor[rgb]{0.902,0.400,0.498}0.164 & \cellcolor[rgb]{0.929,0.400,0.471}0.119 & \cellcolor[rgb]{0.994,0.400,0.406}0.010 \\
Pride & \cellcolor[rgb]{0.839,0.400,0.561}\underline{0.268} & \cellcolor[rgb]{0.833,0.400,0.567}\textbf{0.279} & \cellcolor[rgb]{0.886,0.400,0.514}0.190 & \cellcolor[rgb]{0.893,0.400,0.507}0.179 & \cellcolor[rgb]{1.000,0.400,0.400}$-$0.024 & \cellcolor[rgb]{0.966,0.400,0.434}0.057 & \cellcolor[rgb]{0.897,0.400,0.503}0.171 & \cellcolor[rgb]{0.935,0.400,0.465}0.108 & \cellcolor[rgb]{0.958,0.400,0.442}0.070 & \cellcolor[rgb]{0.913,0.400,0.487}0.145 & \cellcolor[rgb]{1.000,0.400,0.400}$-$0.029 \\
Affection & \cellcolor[rgb]{0.686,0.400,0.714}\textbf{0.523} & \cellcolor[rgb]{0.699,0.400,0.701}\underline{0.501} & \cellcolor[rgb]{0.723,0.400,0.677}0.462 & \cellcolor[rgb]{0.716,0.400,0.684}0.473 & \cellcolor[rgb]{1.000,0.400,0.400}$-$0.047 & \cellcolor[rgb]{1.000,0.400,0.400}$-$0.069 & \cellcolor[rgb]{0.935,0.400,0.465}0.109 & \cellcolor[rgb]{1.000,0.400,0.400}$-$0.024 & \cellcolor[rgb]{1.000,0.400,0.400}$-$0.113 & \cellcolor[rgb]{1.000,0.400,0.400}$-$0.259 & \cellcolor[rgb]{1.000,0.400,0.400}$-$0.157 \\
Pleasure/Ecstasy & \cellcolor[rgb]{0.753,0.400,0.647}\underline{0.411} & \cellcolor[rgb]{0.690,0.400,0.710}\textbf{0.516} & \cellcolor[rgb]{0.821,0.400,0.579}0.298 & \cellcolor[rgb]{0.761,0.400,0.639}0.399 & \cellcolor[rgb]{0.963,0.400,0.437}0.062 & \cellcolor[rgb]{1.000,0.400,0.400}$-$0.388 & \cellcolor[rgb]{0.883,0.400,0.517}0.195 & \cellcolor[rgb]{0.959,0.400,0.441}0.069 & \cellcolor[rgb]{1.000,0.400,0.400}$-$0.096 & \cellcolor[rgb]{1.000,0.400,0.400}$-$0.059 & \cellcolor[rgb]{1.000,0.400,0.400}$-$0.037 \\
Pain & \cellcolor[rgb]{0.785,0.400,0.615}\underline{0.358} & \cellcolor[rgb]{0.782,0.400,0.618}\textbf{0.364} & \cellcolor[rgb]{0.810,0.400,0.590}0.316 & \cellcolor[rgb]{0.804,0.400,0.596}0.326 & \cellcolor[rgb]{1.000,0.400,0.400}$-$0.164 & \cellcolor[rgb]{0.948,0.400,0.452}0.087 & \cellcolor[rgb]{0.931,0.400,0.469}0.115 & \cellcolor[rgb]{0.971,0.400,0.429}0.048 & \cellcolor[rgb]{0.929,0.400,0.471}0.118 & \cellcolor[rgb]{1.000,0.400,0.400}$-$0.043 & \cellcolor[rgb]{1.000,0.400,0.400}$-$0.194 \\
Awe & \cellcolor[rgb]{0.740,0.400,0.660}0.434 & \cellcolor[rgb]{0.717,0.400,0.683}\textbf{0.472} & \cellcolor[rgb]{0.738,0.400,0.662}\underline{0.436} & \cellcolor[rgb]{0.773,0.400,0.627}0.378 & \cellcolor[rgb]{1.000,0.400,0.400}$-$0.143 & \cellcolor[rgb]{0.942,0.400,0.458}0.097 & \cellcolor[rgb]{1.000,0.400,0.400}$-$0.112 & \cellcolor[rgb]{1.000,0.400,0.400}$-$0.159 & \cellcolor[rgb]{1.000,0.400,0.400}$-$0.025 & \cellcolor[rgb]{1.000,0.400,0.400}$-$0.134 & \cellcolor[rgb]{1.000,0.400,0.400}$-$0.054 \\
Intoxication/Altered States & \cellcolor[rgb]{0.893,0.400,0.507}0.178 & \cellcolor[rgb]{0.854,0.400,0.546}\textbf{0.243} & \cellcolor[rgb]{0.857,0.400,0.543}\underline{0.238} & \cellcolor[rgb]{0.864,0.400,0.536}0.227 & \cellcolor[rgb]{1.000,0.400,0.400}$-$0.008 & \cellcolor[rgb]{1.000,0.400,0.400}$-$0.177 & \cellcolor[rgb]{0.943,0.400,0.457}0.095 & \cellcolor[rgb]{0.914,0.400,0.486}0.144 & \cellcolor[rgb]{0.909,0.400,0.491}0.151 & \cellcolor[rgb]{0.963,0.400,0.437}0.061 & \cellcolor[rgb]{0.990,0.400,0.410}0.017 \\
Longing & \cellcolor[rgb]{0.727,0.400,0.673}\textbf{0.455} & \cellcolor[rgb]{0.760,0.400,0.640}\underline{0.400} & \cellcolor[rgb]{0.778,0.400,0.622}0.370 & \cellcolor[rgb]{0.806,0.400,0.594}0.324 & \cellcolor[rgb]{0.987,0.400,0.413}0.022 & \cellcolor[rgb]{1.000,0.400,0.400}$-$0.218 & \cellcolor[rgb]{1.000,0.400,0.400}$-$0.014 & \cellcolor[rgb]{1.000,0.400,0.400}$-$0.028 & \cellcolor[rgb]{1.000,0.400,0.400}$-$0.067 & \cellcolor[rgb]{1.000,0.400,0.400}$-$0.140 & \cellcolor[rgb]{0.985,0.400,0.415}0.025 \\
Concentration & \cellcolor[rgb]{0.837,0.400,0.563}\textbf{0.271} & \cellcolor[rgb]{0.944,0.400,0.456}0.094 & \cellcolor[rgb]{1.000,0.400,0.400}$-$0.041 & \cellcolor[rgb]{1.000,0.400,0.400}$-$0.013 & \cellcolor[rgb]{0.927,0.400,0.473}0.122 & \cellcolor[rgb]{0.859,0.400,0.541}\underline{0.235} & \cellcolor[rgb]{0.945,0.400,0.455}0.092 & \cellcolor[rgb]{0.943,0.400,0.457}0.095 & \cellcolor[rgb]{1.000,0.400,0.400}$-$0.175 & \cellcolor[rgb]{0.863,0.400,0.537}0.229 & \cellcolor[rgb]{0.963,0.400,0.437}0.061 \\
Elation & \cellcolor[rgb]{0.792,0.400,0.608}0.347 & \cellcolor[rgb]{0.755,0.400,0.645}\textbf{0.408} & \cellcolor[rgb]{0.756,0.400,0.644}0.406 & \cellcolor[rgb]{0.756,0.400,0.644}\underline{0.407} & \cellcolor[rgb]{0.918,0.400,0.482}0.136 & \cellcolor[rgb]{1.000,0.400,0.400}$-$0.372 & \cellcolor[rgb]{1.000,0.400,0.400}$-$0.205 & \cellcolor[rgb]{0.918,0.400,0.482}0.136 & \cellcolor[rgb]{1.000,0.400,0.400}$-$0.290 & \cellcolor[rgb]{1.000,0.400,0.400}$-$0.224 & \cellcolor[rgb]{0.906,0.400,0.494}0.157 \\
Doubt & \cellcolor[rgb]{1.000,0.400,0.400}$-$0.038 & \cellcolor[rgb]{0.921,0.400,0.479}0.132 & \cellcolor[rgb]{0.972,0.400,0.428}0.047 & \cellcolor[rgb]{0.962,0.400,0.438}0.064 & \cellcolor[rgb]{0.918,0.400,0.482}0.136 & \cellcolor[rgb]{1.000,0.400,0.400}$-$0.051 & \cellcolor[rgb]{1.000,0.400,0.400}$-$0.026 & \cellcolor[rgb]{0.915,0.400,0.485}0.141 & \cellcolor[rgb]{0.828,0.400,0.572}\textbf{0.287} & \cellcolor[rgb]{0.899,0.400,0.501}\underline{0.169} & \cellcolor[rgb]{0.992,0.400,0.408}0.014 \\
Sourness & \cellcolor[rgb]{1.000,0.400,0.400}$-$0.028 & \cellcolor[rgb]{0.888,0.400,0.512}\underline{0.187} & \cellcolor[rgb]{0.944,0.400,0.456}0.093 & \cellcolor[rgb]{0.911,0.400,0.489}0.149 & \cellcolor[rgb]{0.971,0.400,0.429}0.048 & \cellcolor[rgb]{0.969,0.400,0.431}0.051 & \cellcolor[rgb]{0.932,0.400,0.468}0.113 & \cellcolor[rgb]{1.000,0.400,0.400}$-$0.060 & \cellcolor[rgb]{0.972,0.400,0.428}0.047 & \cellcolor[rgb]{0.880,0.400,0.520}\textbf{0.200} & \cellcolor[rgb]{0.990,0.400,0.410}0.017 \\
Fear & \cellcolor[rgb]{0.880,0.400,0.520}\underline{0.200} & \cellcolor[rgb]{0.849,0.400,0.551}\textbf{0.251} & \cellcolor[rgb]{0.920,0.400,0.480}0.133 & \cellcolor[rgb]{0.901,0.400,0.499}0.165 & \cellcolor[rgb]{1.000,0.400,0.400}$-$0.095 & \cellcolor[rgb]{0.939,0.400,0.461}0.102 & \cellcolor[rgb]{0.978,0.400,0.422}0.037 & \cellcolor[rgb]{1.000,0.400,0.400}$-$0.071 & \cellcolor[rgb]{1.000,0.400,0.400}$-$0.060 & \cellcolor[rgb]{0.923,0.400,0.477}0.128 & \cellcolor[rgb]{1.000,0.400,0.400}$-$0.044 \\
Contentment & \cellcolor[rgb]{0.823,0.400,0.577}0.295 & \cellcolor[rgb]{0.759,0.400,0.641}\textbf{0.401} & \cellcolor[rgb]{0.828,0.400,0.572}0.286 & \cellcolor[rgb]{0.817,0.400,0.583}\underline{0.305} & \cellcolor[rgb]{0.990,0.400,0.410}0.017 & \cellcolor[rgb]{0.959,0.400,0.441}0.068 & \cellcolor[rgb]{1.000,0.400,0.400}$-$0.120 & \cellcolor[rgb]{1.000,0.400,0.400}$-$0.239 & \cellcolor[rgb]{1.000,0.400,0.400}$-$0.285 & \cellcolor[rgb]{1.000,0.400,0.400}$-$0.181 & \cellcolor[rgb]{0.903,0.400,0.497}0.161 \\
Hope/Enthusiasm/Optimism & \cellcolor[rgb]{0.847,0.400,0.553}0.255 & \cellcolor[rgb]{0.842,0.400,0.558}\underline{0.263} & \cellcolor[rgb]{0.839,0.400,0.561}\textbf{0.268} & \cellcolor[rgb]{0.867,0.400,0.533}0.221 & \cellcolor[rgb]{1.000,0.400,0.400}$-$0.156 & \cellcolor[rgb]{1.000,0.400,0.400}$-$0.157 & \cellcolor[rgb]{1.000,0.400,0.400}$-$0.158 & \cellcolor[rgb]{0.982,0.400,0.418}0.030 & \cellcolor[rgb]{1.000,0.400,0.400}$-$0.042 & \cellcolor[rgb]{1.000,0.400,0.400}$-$0.023 & \cellcolor[rgb]{0.891,0.400,0.509}0.181 \\
Infatuation & \cellcolor[rgb]{0.910,0.400,0.490}\underline{0.150} & \cellcolor[rgb]{0.828,0.400,0.572}\textbf{0.286} & \cellcolor[rgb]{0.918,0.400,0.482}0.137 & \cellcolor[rgb]{0.932,0.400,0.468}0.114 & \cellcolor[rgb]{1.000,0.400,0.400}$-$0.071 & \cellcolor[rgb]{1.000,0.400,0.400}$-$0.016 & \cellcolor[rgb]{0.935,0.400,0.465}0.109 & \cellcolor[rgb]{1.000,0.400,0.400}$-$0.041 & \cellcolor[rgb]{1.000,0.400,0.400}$-$0.018 & \cellcolor[rgb]{0.993,0.400,0.407}0.012 & \cellcolor[rgb]{1.000,0.400,0.400}$-$0.022 \\
Interest & \cellcolor[rgb]{0.972,0.400,0.428}0.046 & \cellcolor[rgb]{0.966,0.400,0.434}0.057 & \cellcolor[rgb]{0.926,0.400,0.474}0.124 & \cellcolor[rgb]{0.902,0.400,0.498}\textbf{0.163} & \cellcolor[rgb]{1.000,0.400,0.400}$-$0.105 & \cellcolor[rgb]{0.999,0.400,0.401}0.002 & \cellcolor[rgb]{1.000,0.400,0.400}$-$0.059 & \cellcolor[rgb]{0.944,0.400,0.456}0.094 & \cellcolor[rgb]{0.956,0.400,0.444}0.074 & \cellcolor[rgb]{0.908,0.400,0.492}\underline{0.154} & \cellcolor[rgb]{0.978,0.400,0.422}0.037 \\
Shame & \cellcolor[rgb]{0.998,0.400,0.402}0.004 & \cellcolor[rgb]{0.829,0.400,0.571}\textbf{0.285} & \cellcolor[rgb]{0.851,0.400,0.549}0.249 & \cellcolor[rgb]{0.844,0.400,0.556}\underline{0.260} & \cellcolor[rgb]{0.967,0.400,0.433}0.055 & \cellcolor[rgb]{1.000,0.400,0.400}$-$0.127 & \cellcolor[rgb]{1.000,0.400,0.400}$-$0.028 & \cellcolor[rgb]{0.927,0.400,0.473}0.122 & \cellcolor[rgb]{1.000,0.400,0.400}$-$0.147 & \cellcolor[rgb]{1.000,0.400,0.400}$-$0.043 & \cellcolor[rgb]{1.000,0.400,0.400}$-$0.122 \\
Jealousy/Envy & \cellcolor[rgb]{0.918,0.400,0.482}0.137 & \cellcolor[rgb]{0.857,0.400,0.543}\textbf{0.239} & \cellcolor[rgb]{0.944,0.400,0.456}0.093 & \cellcolor[rgb]{0.899,0.400,0.501}\underline{0.168} & \cellcolor[rgb]{0.945,0.400,0.455}0.091 & \cellcolor[rgb]{1.000,0.400,0.400}$-$0.079 & \cellcolor[rgb]{1.000,0.400,0.400}$-$0.064 & \cellcolor[rgb]{1.000,0.400,0.400}$-$0.093 & \cellcolor[rgb]{0.955,0.400,0.445}0.075 & \cellcolor[rgb]{1.000,0.400,0.400}$-$0.067 & \cellcolor[rgb]{1.000,0.400,0.400}$-$0.095 \\
Emotional Numbness & \cellcolor[rgb]{0.725,0.400,0.675}0.459 & \cellcolor[rgb]{0.714,0.400,0.686}\textbf{0.476} & \cellcolor[rgb]{1.000,0.400,0.400}$-$0.233 & \cellcolor[rgb]{1.000,0.400,0.400}$-$0.198 & \cellcolor[rgb]{0.807,0.400,0.593}0.322 & \cellcolor[rgb]{0.717,0.400,0.683}\underline{0.472} & \cellcolor[rgb]{0.860,0.400,0.540}0.234 & \cellcolor[rgb]{1.000,0.400,0.400}$-$0.430 & \cellcolor[rgb]{1.000,0.400,0.400}$-$0.180 & \cellcolor[rgb]{1.000,0.400,0.400}$-$0.278 & \cellcolor[rgb]{1.000,0.400,0.400}$-$0.300 \\
Sexual Lust & \cellcolor[rgb]{0.935,0.400,0.465}0.109 & \cellcolor[rgb]{0.888,0.400,0.512}\textbf{0.187} & \cellcolor[rgb]{0.914,0.400,0.486}\underline{0.143} & \cellcolor[rgb]{0.923,0.400,0.477}0.128 & \cellcolor[rgb]{1.000,0.400,0.400}$-$0.071 & \cellcolor[rgb]{0.972,0.400,0.428}0.046 & \cellcolor[rgb]{0.979,0.400,0.421}0.035 & \cellcolor[rgb]{1.000,0.400,0.400}$-$0.037 & \cellcolor[rgb]{1.000,0.400,0.400}$-$0.018 & \cellcolor[rgb]{1.000,0.400,0.400}$-$0.076 & \cellcolor[rgb]{1.000,0.400,0.400}$-$0.107 \\
Astonishment/Surprise & \cellcolor[rgb]{0.845,0.400,0.555}\textbf{0.258} & \cellcolor[rgb]{0.880,0.400,0.520}\underline{0.200} & \cellcolor[rgb]{0.990,0.400,0.410}0.017 & \cellcolor[rgb]{1.000,0.400,0.400}$-$0.001 & \cellcolor[rgb]{1.000,0.400,0.400}$-$0.001 & \cellcolor[rgb]{1.000,0.400,0.400}$-$0.189 & \cellcolor[rgb]{0.950,0.400,0.450}0.083 & \cellcolor[rgb]{1.000,0.400,0.400}$-$0.082 & \cellcolor[rgb]{1.000,0.400,0.400}$-$0.105 & \cellcolor[rgb]{1.000,0.400,0.400}$-$0.018 & \cellcolor[rgb]{0.986,0.400,0.414}0.023 \\\bottomrule
\end{tabular}}
\vspace{0.2cm}
\end{table}

\subsection{\voicenetext Per-Attribute \texorpdfstring{$\rho$}{rho}}
\label{app:dim_per_attribute}

Per-attribute Spearman $\rho$ for \voicenetext covers 57 attributes $\times$ 7 levels (399 (attribute, level) prompts). The full per-attribute $\rho$ table is released alongside the paper rather than reproduced here, because the averages quoted in the main text ($\rho{=}0.15$ for \voiceclaplarge, $\rho{=}0.11$ for \voiceclapsmall) summarise the signal adequately: per-attribute values are noisy at $\sim$46 audios per (attribute, level) prompt. We additionally release \texttt{per\_attribute\_reliability.csv}, containing one row per attribute with total and per-rater-count coverage, majority-yes prevalence and Wilson interval, strict three-rater observed agreement and Fleiss $\kappa$, item-bootstrap 95\% intervals, and reliable-core membership. Its aggregate exact estimate is $\kappa=-0.0218$ with a 95\% item-bootstrap interval $[-0.0367,-0.0066]$, which excludes zero. The accompanying script fixes the bootstrap seed and records the source-annotation SHA-256. Tab.~\ref{tab:dim_per_attribute_coverage} records the underlying coverage that each released $\rho$ value is computed from --- the per-attribute pair count $n$ and the human-vote distribution --- so readers can see at a glance which attributes are well-sampled (e.g.\ \texttt{S\_RANT}, \texttt{ARSH}, \texttt{S\_DRAM} at $n{=}350$) and which are sparser (\texttt{FULL} at 273 pairs, \texttt{EXPL} at 144).

\begin{table}[t]
\centering
\scriptsize
\setlength{\tabcolsep}{4pt}
\renewcommand{\arraystretch}{1.0}
\caption{Per-attribute coverage and majority-vote statistics on \voicenetext. $n$ is the number of (audio, attribute-level) pairs released for that attribute, aggregated across the seven rubric levels; per-(attribute, level) prompt counts are therefore roughly $n/7$. \emph{maj-yes} is the fraction of items where the rater majority voted ``present''; \emph{polar.~match} is the fraction of items where the majority vote agrees with the prompt's intended affirmative-vs.-contrastive polarity. Per-attribute Spearman $\rho$ for each model is computed over the same $n$ items per attribute and is released alongside the paper as a data artifact. Attribute codes follow the schema in App.~\ref{app:dim_attributes}; rows are ordered to match that grouping.}
\label{tab:dim_per_attribute_coverage}
\begin{tabular}{@{}llrcc@{}}
\toprule
Code & Description & $n$ & maj-yes & polar.~match \\
\midrule
\multicolumn{5}{@{}l}{\textit{Perceived speaker (2)}} \\
\texttt{GEND}    & gender presentation       & 350 & 0.237 & 0.594 \\
\texttt{AGEV}    & age range                 & 349 & 0.284 & 0.564 \\
\multicolumn{5}{@{}l}{\textit{Affective dimensions (7)}} \\
\texttt{VALN}    & valence                   & 343 & 0.315 & 0.624 \\
\texttt{AROU}    & arousal                   & 343 & 0.280 & 0.618 \\
\texttt{VOLT}    & volatility                & 318 & 0.343 & 0.566 \\
\texttt{VALS}    & valence sharpness         & 328 & 0.216 & 0.573 \\
\texttt{TENS}    & vocal tension             & 302 & 0.288 & 0.583 \\
\texttt{ARSH}    & acoustic harshness        & 350 & 0.206 & 0.574 \\
\texttt{VULN}    & vulnerability             & 350 & 0.294 & 0.571 \\
\multicolumn{5}{@{}l}{\textit{Prosodic delivery (12)}} \\
\texttt{TEMP}    & speaking rate             & 318 & 0.236 & 0.531 \\
\texttt{ATCK}    & attack                    & 350 & 0.357 & 0.623 \\
\texttt{CHNK}    & chunking                  & 321 & 0.280 & 0.523 \\
\texttt{RANG}    & melodic range             & 302 & 0.288 & 0.540 \\
\texttt{VFLX}    & vocal flexibility         & 350 & 0.174 & 0.531 \\
\texttt{STNC}    & vocal stance              & 342 & 0.310 & 0.640 \\
\texttt{EMPH}    & emphasis                  & 313 & 0.339 & 0.607 \\
\texttt{DFLU}    & disfluency                & 340 & 0.332 & 0.647 \\
\texttt{CLRT}    & clarity                   & 332 & 0.310 & 0.572 \\
\texttt{STRU}    & discourse structure       & 305 & 0.334 & 0.626 \\
\texttt{COGL}    & cognitive load            & 320 & 0.328 & 0.600 \\
\texttt{FOCS}    & focus                     & 341 & 0.238 & 0.560 \\
\multicolumn{5}{@{}l}{\textit{Vocal quality (12)}} \\
\texttt{ROUG}    & roughness                 & 303 & 0.267 & 0.587 \\
\texttt{SMTH}    & smoothness                & 314 & 0.287 & 0.564 \\
\texttt{BRGT}    & brightness                & 320 & 0.291 & 0.569 \\
\texttt{WARM}    & warmth                    & 314 & 0.255 & 0.599 \\
\texttt{FULL}    & fullness                  & 273 & 0.245 & 0.560 \\
\texttt{HARM}    & harmonicity               & 313 & 0.198 & 0.556 \\
\texttt{METL}    & metalicity                & 322 & 0.261 & 0.581 \\
\texttt{ESTH}    & esthetics                 & 317 & 0.322 & 0.590 \\
\texttt{REGS}    & register                  & 350 & 0.211 & 0.574 \\
\texttt{RESP}    & respiration               & 344 & 0.334 & 0.616 \\
\texttt{DARC}    & dark/light                & 329 & 0.216 & 0.568 \\
\texttt{RCQL}    & recording quality         & 286 & 0.329 & 0.451 \\
\multicolumn{5}{@{}l}{\textit{Resonance placement (7)}} \\
\texttt{R\_CHST} & chest                     & 320 & 0.316 & 0.537 \\
\texttt{R\_THRT} & throat                    & 336 & 0.271 & 0.583 \\
\texttt{R\_ORAL} & oral                      & 311 & 0.280 & 0.566 \\
\texttt{R\_HEAD} & head                      & 340 & 0.271 & 0.503 \\
\texttt{R\_MASK} & mask                      & 332 & 0.304 & 0.596 \\
\texttt{R\_MIXD} & mixed                     & 318 & 0.252 & 0.560 \\
\texttt{R\_NASL} & nasal                     & 279 & 0.226 & 0.595 \\
\multicolumn{5}{@{}l}{\textit{Recording context (2)}} \\
\texttt{BKGN}    & background noise          & 315 & 0.273 & 0.546 \\
\texttt{EXPL}    & expletive content         & 144 & 0.361 & 0.660 \\
\multicolumn{5}{@{}l}{\textit{Style descriptors (15)}} \\
\texttt{S\_CONV} & conversational            & 333 & 0.288 & 0.643 \\
\texttt{S\_CASU} & casual                    & 316 & 0.282 & 0.620 \\
\texttt{S\_PLAY} & playful                   & 350 & 0.254 & 0.594 \\
\texttt{S\_CART} & cartoonish                & 347 & 0.303 & 0.631 \\
\texttt{S\_FORM} & formal                    & 342 & 0.319 & 0.643 \\
\texttt{S\_AUTH} & authoritative             & 338 & 0.302 & 0.660 \\
\texttt{S\_TECH} & teacher/didactic          & 322 & 0.323 & 0.575 \\
\texttt{S\_MONO} & monologue                 & 350 & 0.371 & 0.609 \\
\texttt{S\_DRAM} & dramatic                  & 350 & 0.260 & 0.640 \\
\texttt{S\_NARR} & narrator                  & 329 & 0.307 & 0.614 \\
\texttt{S\_STRY} & storytelling              & 335 & 0.346 & 0.648 \\
\texttt{S\_NEWS} & newsreader                & 340 & 0.315 & 0.647 \\
\texttt{S\_RANT} & ranting                   & 350 & 0.266 & 0.640 \\
\texttt{S\_WHIS} & whisper-talk              & 350 & 0.300 & 0.651 \\
\texttt{S\_ASMR} & ASMR                      & 333 & 0.255 & 0.625 \\
\midrule
\textbf{Total / mean} & \textbf{57 attributes} & \textbf{18{,}532} & \textbf{0.284} & \textbf{0.591} \\
\bottomrule
\end{tabular}
\end{table}

\subsection{Human Audit of the MOSS-Audio Labels (Interim)}
\label{app:moss_audit}

Because the dense \textsc{MOSS-Audio} annotations that supervise \voiceclap are model-generated, we run a prespecified audit of how closely a human rater reproduces them. The design was fixed in advance: 300 items, 75 English clips drawn from each of the four \textsc{MOSS}-annotated corpora, one attribute per clip, all 59 prompt-group attributes represented, plus 24 hidden checks that pair a clip with a genuine \textsc{MOSS} sentence written for a different clip at a distant level of the same attribute. The rater first selects a rubric level independently, and only then is the candidate \textsc{MOSS} label revealed and judged for fit, so the exact-level statistic cannot be anchored on the machine's answer. The sample is English-only so that every attribute is answerable and comprehension never confounds the judgement; it therefore measures annotation quality on the English portion of corpora that ship multilingual. Since the corpora store one free-text sentence per (clip, attribute) and no numeric level, each sentence was mapped back onto its rubric by a text-only pass run with the anchors in both directions, and only items on which both directions agree carry a level and enter the level-match statistic.

The audit is not finished. Five raters have contributed 310 judgements covering 97 of the 300 items, and no rater has completed a full pass, so intervals resample items rather than judgements and the result below is interim. Raters chose the same rubric level as \textsc{MOSS} on $0.257$ of mapped judgements (95\% CI $[0.191, 0.325]$) against a permutation baseline of $0.169$, and within one level on $0.563$ (95\% CI $[0.485, 0.646]$) against $0.473$. Because several raters judged the same items, the identical statistic is computable between raters: two raters chose the same level on $0.275$ (95\% CI $[0.220, 0.337]$). The paired difference between human--human and human--\textsc{MOSS} agreement is $+0.018$ (95\% CI $[-0.053, +0.102]$) for exact match and $+0.087$ (95\% CI $[-0.010, +0.192]$) within one level; both contain zero. The rank correlation between the \textsc{MOSS} level and the rater level is $0.237$, which falls between the 50th and 75th percentile of the effect-size scale discussed in \S\ref{sec:discussion} \citep{gignac2016effectsize}.

Two properties of the instrument belong alongside those numbers. Raters were also asked whether the \textsc{MOSS} sentence fits the clip and accepted it in $0.732$ of judgements, but only 2 of 12 hidden checks were rejected, while on those checks their own level choice sat $2.2$ levels from the planted label against $1.5$ on real items. Raters located the discrepancy and then declined to call it a mismatch, so the fit judgement is too permissive to serve as the primary measure and exact level match is used instead. The sample supports overall statements only, not per-attribute ones. The audit set, the per-item judgements, and the analysis script are available at \url{https://github.com/LAION-AI/emolia-bench}; the completed audit will be reported whatever it shows.

\subsection{Detailed Taxonomy Construction Methodology}
\label{app:taxonomy_construction}

The 40-category emotion taxonomy utilized in both the \emonet foundation and benchmark datasets was originally developed for the EmoNet-Face Benchmark \citep{schuhmann2025emonetface}.

The taxonomy aims to support fine-grained understanding of affective states beyond basic-emotion models, drawing on the Theory of Constructed Emotion (TCE)~\citep{barrett2017theory}.

The taxonomy spans positive and negative emotions, social emotions (\textit{Embarrassment}, \textit{Shame}, and \textit{Pride}), cognitive states (\textit{Concentration}, \textit{Doubt}, and \textit{Confusion}), bodily states (\textit{Pain}, \textit{Fatigue}, and \textit{Intoxication}), and less common categories such as \textit{Sourness} and \textit{Helplessness}. The full list of 40 categories with descriptive word clusters appears in App.~\ref{app:taxonomy}; the same list is used in the prior EmoNet-Face paper.

The construction process involved several key stages:
\begin{enumerate}
    \item \textbf{Literature-Driven Candidate Extraction:} The comprehensive "Handbook of Emotions" (946 pages)~\citep{lewis2016handbook} was digitized using Optical Character Recognition (OCR). The digitized text was then divided into manageable 500-word segments.
    \item \textbf{AI-Assisted Term Identification:} GPT-4 was employed to analyze these text segments and extract potential nouns representing emotion concepts.
    \item \textbf{Refinement and Deduplication:} The initially extracted terms were aggregated, and duplicates were removed, resulting in a candidate list of approximately 170 unique emotion-related nouns.
    \item \textbf{Expert-Guided Clustering and Categorization:} This refined list of terms underwent an iterative process of clustering. This involved independent categorization efforts by the taxonomy authors, followed by critical reviews and discussions. Psychologists and researchers in affective computing provided expert guidance throughout this phase to ensure the semantic coherence and psychological relevance of the emerging categories. Each of the final 40 categories represents a cluster of these semantically related emotion words.
\end{enumerate}

In line with the Theory of Constructed Emotion, this taxonomy does not presuppose the biological universality or fixedness of these emotional categories. Instead, it is intended to facilitate context-aware and socially informed interpretations of affective expressions by AI systems. Recognizing the inherent ambiguity in perceiving emotions (e.g., a high-arousal vocal expression might be interpreted as amusement, elation, or excitement depending on context and observer), the taxonomy was specifically designed to support plausible multi-label annotations rather than forcing rigid, single-label classifications. This approach aims to enable richer and more contextually sensitive representations of emotion in AI.

\subsection{Per-Emotion Annotator Agreement on \voicenetemo}
\label{app:annotator_agreement}

\begin{figure}[t]
    \centering
    \includegraphics[width=\textwidth]{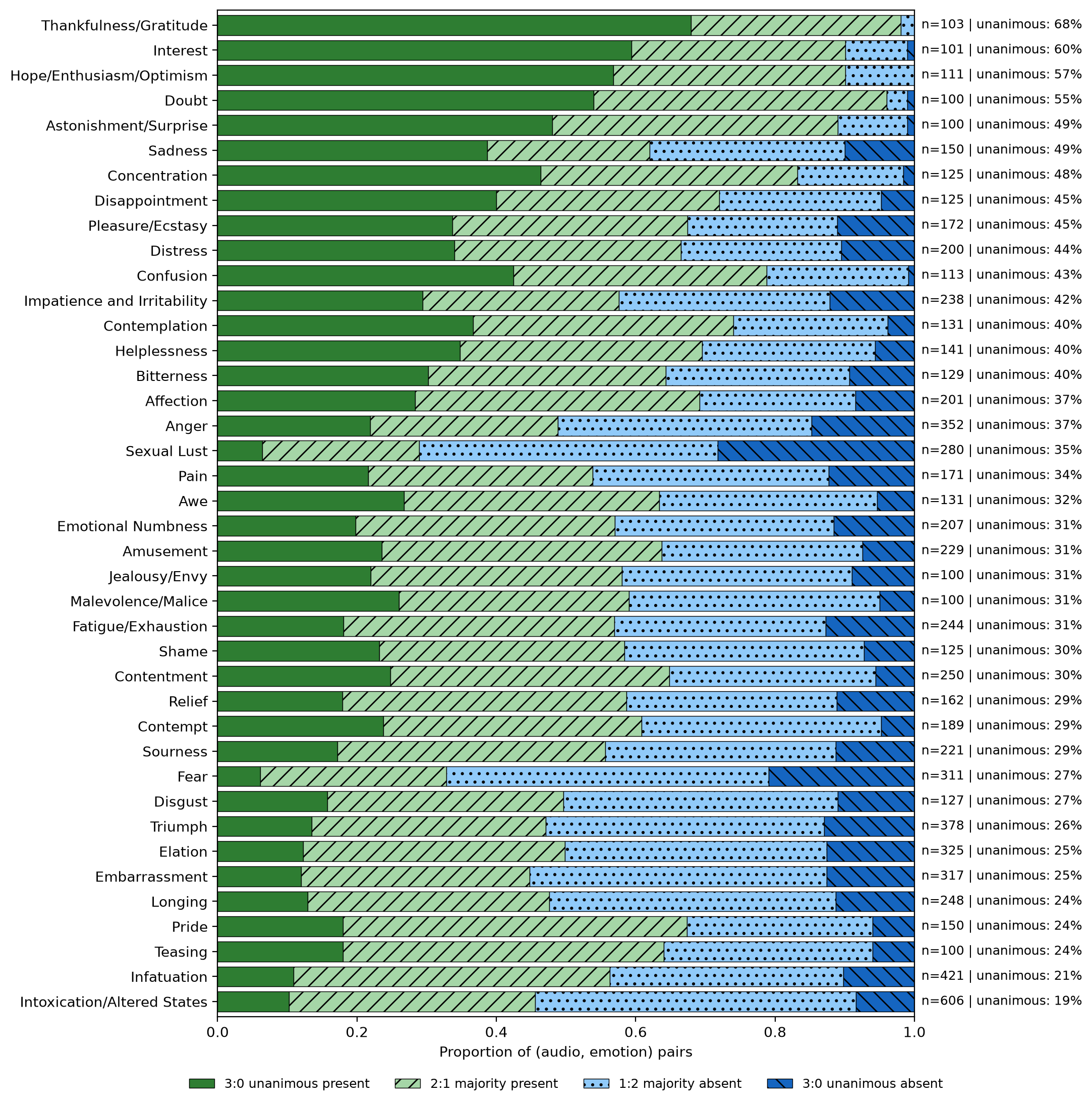}
    \caption{Expert agreement on emotion presence for \voicenetemo, per emotion, on the 7,984 pairs with three ratings. Stacked bars show the proportion of (audio, emotion) pairs by the number of experts who marked the emotion present (weakly or strongly): three (\textsf{3:0 unanimous present}), two (\textsf{2:1}), one (\textsf{1:2}), or none (\textsf{3:0 unanimous absent}). The text to the right of each bar gives the number of pairs $n$ and the share of unanimous pairs. Rows are sorted by unanimity.}
    \label{fig:annotator_agreement_detailed_human}
\end{figure}

Fig.~\ref{fig:annotator_agreement_detailed_human} reports the distribution of binary (present vs.\ absent) three-rater outcomes per emotion on the 7,984 fully rated \voicenetemo pairs. For each emotion, the stacked bar shows the fraction of pairs on which all three experts mark the emotion present (\textsf{3:0 unanimous present}), two of three mark it present (\textsf{2:1}), one of three marks it present (\textsf{1:2}), or none does (\textsf{3:0 unanimous absent}). Across all pairs, the three experts are unanimous on 32.7\% of items. Unanimity is highest for \emph{Thankfulness/Gratitude} (68\%), \emph{Interest} (60\%), \emph{Hope/Enthusiasm/Optimism} (57\%), and \emph{Doubt} (55\%), almost entirely through unanimous-present votes, and lowest for \emph{Intoxication/Altered States} (19\%), \emph{Infatuation} (21\%), and \emph{Teasing}, \emph{Pride}, and \emph{Longing} (24\% each). Per-emotion unanimity is unrelated to per-emotion model performance in Tab.~\ref{tab:model_results_by_emotion}: across the 40 emotions its Spearman correlation with \voiceclaplarge $\rho$ is $0.02$, and \emph{Interest} and \emph{Doubt} combine high unanimity with some of the weakest model signal. This matches the agreement-stratified analysis in \S\ref{sec:per_attribute}, which finds no concentration of model gains in higher-agreement emotions.

\subsection{\voiceclap Training Details}
\label{app:training_details}

\textbf{\voiceclapsmall.} A 110M-parameter dual-tower CLAP \citep{wu2023clap}: BUD-E-Whisper-Small \citep{radford2023whisper} (768-d audio output) and \texttt{all-MiniLM-L6-v2} \citep{reimers2019sentence,wang2020minilm} with mean-pooling (384-d text output), each with a learnable linear projection to a shared 768-d space. Training uses SigLIP sigmoid contrastive loss \citep{zhai2023sigmoid}, AdamW ($\beta_1{=}0.9$, $\beta_2{=}0.95$, weight decay 0.2), LR $10^{-4}$ with 200-step warmup and cosine decay, gradient clip 1.0, per-GPU batch 128 on 4 GH200s (effective batch 512 with all-gather), one epoch ($\sim$0.5h).

\textbf{\voiceclaplarge.} A single-tower rank-16 LoRA \citep{hu2022lora} (alpha 32, dropout 0.05) finetune of \textsc{LCO-Embedding-Omni-7B} \citep{xiao2025scaling} (Qwen2.5-Omni-Thinker-7B backbone with sentence-transformer head, 3,584-d output). Training uses symmetric InfoNCE \citep{oord2018representation} on cosine similarities at fixed logit scale $\log(1/0.07)$ with all-gather negatives, AdamW ($\beta_2{=}0.95$, weight decay 0.01), LR $10^{-4}$ (200-step warmup, cosine decay), gradient clip 1.0. On four GH200s, micro-batch 2 with 16-step accumulation gives a 128 optimizer-step batch and an effective contrastive batch of 8; accumulation reduces noise but does not enlarge the InfoNCE negative pool. One epoch ($\sim$2h).

\subsection{\voicenetext Annotation Coverage and Reliability}
\label{app:ext_reliability}

\voicenetext asks: \emph{does clip $C$ exhibit attribute $A$ at level $L$?} for the 57 talking-style attributes of \S\ref{sec:taxonomy}. The audio source is distinct from \emoliabal. Each of the 57 dimensions is defined by a rubric with ordinal levels (seven levels, 0--6, for nearly all attributes); the full rubric definitions are available at \url{https://projects.laion.ai/emolia-bench/taxonomy/}. For each (dimension, level) bucket, the annotation task is reduced to a binary confirmation: candidate audio clips are pre-screened by Gemini~3~Flash, which confirms whether each clip matches the target level description. An equal number of positives (clips confirmed to match the target level) and negatives (clips confirmed to match a non-adjacent level of the same dimension) are presented to human annotators, who listen and confirm or reject the match. This design ensures that annotators perform a focused binary judgement rather than a full ordinal rating. The release contains 18.5k (audio, attribute-level) pairs carrying 40{,}990 ratings from eight annotators; four hold psychology degrees (BSc to MSc, including clinical psychology) and contributed 95.3\% of all ratings, while the remaining four come from law, translation, computer science, and environmental-science backgrounds. 5{,}828 pairs carry at least three ratings (5{,}583 with exactly three, 241 with four, and 4 with five), 10{,}338 carry two, and 2{,}366 carry one.
Aggregate majority-yes rate is 0.284. On the 5{,}583 non-\texttt{LANG} items with exactly three ratings, Fleiss' binary $\kappa=-0.022$ (95\% item-bootstrap CI $[-0.037,-0.007]$) and observed rater-pair agreement is 0.494 $[0.486,0.501]$; pairwise and leave-one-out balanced-accuracy references are 0.533 and 0.479 (the pairwise figure averages over rater pairs; weighting instead by the 28{,}573 paired votes gives 0.484). With 5{,}583 three-rater items the $\kappa$ interval excludes zero, so near-chance agreement on \voicenetext is a stable property of these fine-grained rubric-level judgements as currently operationalised, not an artefact of sparse annotation. We prespecified a reliable-core rule before computing the per-attribute results: at least 20 three-rater items, $\kappa\geq0.20$, and a 95\% lower bound above zero. Every attribute clears the sample-size bar (all 57 have $n\geq20$ strict three-rater items, median $n{=}98$), but none reaches $\kappa\geq0.20$: the maximum is \texttt{STNC} at $\kappa{=}0.085$ $[-0.043,0.211]$, only 17 of 57 attributes have $\kappa>0$, and the median is $-0.031$. No attribute therefore enters the uncertainty-supported core. We therefore label all current \voicenetext scores preliminary and treat its attributes as exploratory perceptual probes. Per-attribute prevalence, coverage, agreement, $\kappa$, uncertainty, and model $\rho$ files are released alongside the benchmark (App.~\ref{app:dim_per_attribute}).

\subsection{Additional Results}
\label{app:additional_results}

\textbf{Human baselines and ceilings.} Four reference points anchor the model numbers. Random and always-predict-majority both give 0.500 balanced accuracy by construction (with always-predict-majority reaching 0.578 raw accuracy on \voicenetemo and 0.716 on \voicenetext due to the present/absent class skew). The pairwise human ceiling scores rater A's binary vote against rater B's on items with $\geq 2$ raters: 0.562 balanced accuracy on \voicenetemo (24.0k paired votes) and 0.533 on \voicenetext (averaged over rater pairs; 0.484 if weighted by the 28.6k paired votes instead). The leave-one-out ceiling scores one rater against the majority of the others on items with at least three ratings: 0.572 on \voicenetemo (13.2k comparisons) and 0.479 on \voicenetext (9,246 LOO comparisons across the 5,828 items with three or more ratings). On \voicenetemo, \voiceclaplarge's 0.7021 \textit{bal@pp} is above the 0.572 LOO reference, i.e.\ closer to the multi-rater consensus than any individual expert is. This is not a like-for-like comparison---the model is scored against the aggregate majority label while each expert is scored against the other raters---so it should be read as strong alignment with the expert consensus, not as surpassing human emotion perception. On \voicenetext the human ceilings sit at chance level, below our model scores, reflecting both the harder rubric task and the more heterogeneous rater pool. Human consistency on \voicenetext is therefore at chance on the current rubric, which is why we treat its model scores as exploratory separability rather than a validated ceiling comparison.

\textbf{Calibration.}
Per-prompt thresholding most benefits models whose similarity distribution is biased away from zero. \voiceclaplarge has $\textit{bal@0}=0.580$, $\textit{bal@opt}=0.673$ at $\theta{=}0.136$, and $\textit{bal@pp}=0.702$. The 12-pp gap from $\textit{bal@0}$ to $\textit{bal@pp}$ reflects a positive similarity bias (mean cosine $\approx 0.14$) and would be largely closed by post-hoc mean-centring. \voiceclapsmall has $\textit{bal@0}=0.659$, $\textit{bal@opt}=0.661$ at $\theta{=}{-}0.020$, and $\textit{bal@pp}=0.675$. Its similarity distribution is well-centred and offers almost no calibration headroom: the bal@pp lead over LCO-Omni-7B (0.6754 vs.\ 0.6632) shrinks to a near-tie under bal@0. The headline ordering between the four families (general-audio CLAPs $<$ Omni bases $\approx$ \voiceclapsmall $<$ \voiceclaplarge) is preserved under bal@opt and per-prompt $\rho$, but readers comparing two models within the same family should consult App.~\ref{app:full_summaries} for both metrics.

\textbf{Agreement-stratified results.} We split the 40 emotions and 57 non-\texttt{LANG} attributes into equal-size lower- and higher-agreement ranks using observed annotator-pair agreement on strict three-rater items, before inspecting model scores. On \voicenetemo, the \voiceclaplarge{} minus \textsc{LCO-Embedding-Omni-7B} advantage is $+0.0436$ versus $+0.0306$ \textit{bal@pp} and $+0.0787$ versus $+0.0548$ $\rho$ in the lower- versus higher-agreement halves. The corresponding high-minus-low interactions are $-0.0130$ (95\% CI $[-0.0350,0.0076]$) and $-0.0239$ $[-0.0628,0.0159]$. On preliminary \voicenetext, the advantages are $+0.0210$ versus $+0.0130$ \textit{bal@pp} and $+0.0453$ versus $+0.0541$ $\rho$; interactions are $-0.0080$ $[-0.0270,0.0121]$ and $+0.0088$ $[-0.0274,0.0481]$. Thus none of the four paired interactions excludes zero: these data do not show that VoiceCLAP's edge is concentrated in the relatively higher-agreement categories. Here ``higher'' is only a within-\voicenetext rank, not evidence that those attributes are reliably annotated.

\textbf{Cross-dataset gap diagnosis.} A paired follow-up analysis clarifies the two columns where \voiceclaplarge trails its 7B base. On synthetic EmoNet-Voice, the same-runtime Large-minus-base gap is $-0.0115$ (95\% clip-bootstrap CI $[-0.0172,-0.0058]$), with gains and losses distributed unevenly across emotion classes rather than a uniform regression. On RAVDESS, the micro-accuracy gap is $-0.0139$, but its actor-clustered interval includes zero ($[-0.0368,0.0104]$); Large simultaneously raises macro F1 from 0.285 to 0.330 and macro recall from 0.300 to 0.348, with the micro loss dominated by a calm/neutral boundary shift. Because IEMOCAP, RAVDESS, and CREMA-D all contain acted speech while Large improves on two of them, these results do not support a generic acted-versus-naturalistic explanation. They instead indicate dataset- and taxonomy-specific boundary changes that micro accuracy alone can hide.

\begin{figure}[h!]
    \centering
    \scriptsize
    \includegraphics{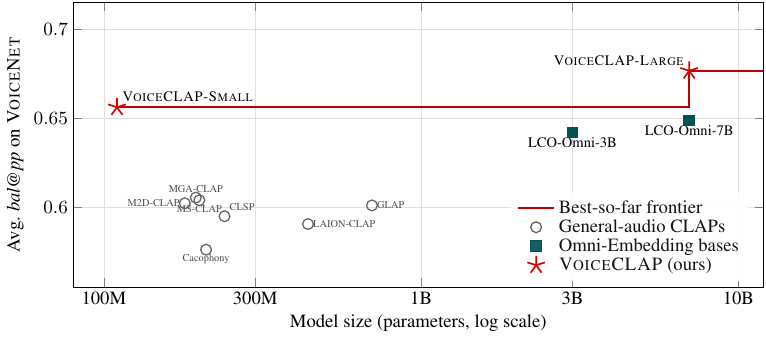}
    \caption{\textbf{Scaling view of \voicenet performance.} Each point is one model. The y-axis is the unweighted mean of \voicenetemo and \voicenetext per-prompt balanced accuracy from Table~\ref{tab:model_results}. The red step line tracks the best score reached by any model up to each parameter count. \voiceclapsmall (110M) sits above the general-audio CLAP cluster and is competitive with the larger Omni bases, while the 7B \voiceclaplarge reaches the highest average score. Parameter counts are approximate (audio + text encoder, before LoRA adapters). Averages match Tab.~\ref{tab:model_results}. }
    \label{fig:scale_frontier}
\end{figure}

\textbf{Scaling view.} Fig.~\ref{fig:scale_frontier} plots the average \textit{bal@pp} of Tab.~\ref{tab:model_results} against model size.

\subsection{Effect Sizes in Psychological Context}
\label{app:effect_sizes}

\textbf{Interpreting the correlations in psychological context.} To contextualise the $\rho$ values in Tab.~\ref{tab:model_results}, we draw on modern benchmarks for effect sizes in psychological research. A large-scale meta-analysis finds that the 25th, 50th, and 75th percentiles for correlations in individual-differences research are $r{=}.11$, $.19$, and $.29$, respectively \citep{gignac2016effectsize}. By these standards, the seven general-audio CLAPs (all $|\rho|<0.1$ on \voicenetemo) fall below even the 25th percentile---they carry essentially no emotion-ranking signal. Notably, this includes CLSP \citep{yang2026towards}, a speech-specialised contrastive model built on the SPEAR-XLarge speech encoder \citep{yang2025spear}, which achieves only $\rho{=}0.061$ on \voicenetemo despite being explicitly designed for speech rather than general audio; speech-aware pretraining alone is insufficient without dense vocal-style supervision. The two Omni-Embedding bases ($\rho{=}0.29$--$0.31$) approach the 75th percentile of published psychological effect sizes. \voiceclaplarge ($\rho{=}0.372$) surpasses this further, and its per-emotion peaks---distress ($0.63$), impatience ($0.58$), anger ($0.58$), sadness ($0.57$)---reach correlations that are exceedingly rare in psychological research.

These numbers must be read against the fundamental limits of \emph{empathic accuracy}---the ability to correctly judge the thoughts and feelings of others---which is ``inherently an interpersonal process that requires active engagement of the emotional systems of both interaction partners'' \citep{lin2024empathic}. Because there is no perfect objective ground truth for perceived emotion intensity, human raters naturally disagree on both category and degree \citep{rum2020empathic,zaki2009empathic}. A model achieving $\rho{=}0.37$ against aggregated expert ratings on a 40-class taxonomy of naturalistic in-the-wild speech has therefore extracted a robust rank-ordered signal from inherently subjective human data. Modern methodological work further cautions against ``heuristically dismissing `small' effects as unimportant'' \citep{anvari2021effects}, since even moderate correlations carry substantial predictive value in complex real-world dynamics. By these standards, \voiceclaplarge's emotion-ranking performance is not merely competitive---it represents a strong effect size that substantially exceeds what is typical in published psychological research.

\subsection{Extended Limitations}
\label{app:limitations}

The evaluation is representation-level throughout. All eleven systems are voice-text embedding models scored by cosine similarity, so the results characterise how well such models rank vocal-attribute descriptions; they do not establish that better \voicenet scores translate into better end-to-end voice assistants, conversational turn-taking, or response generation, and no spoken-dialogue system is evaluated here. Generative audio-language models are likewise out of scope, as prompting them for graded attribute presence requires a scoring interface that is not comparable to the contrastive protocol used here.
\emolia and the three sibling voice corpora carry the source-bias of the open Emilia, LAION's Got Talent, Majestrino, and Multilingual In The Wild collections. Geographic, demographic, and recording-environment coverage are non-uniform. The released 18-prompt-group MOSS-Audio annotations are themselves model-generated; they reflect one model's interpretation of vocal style and are not validated against human ratings at scale. A prespecified human audit of these labels is underway and reported as an interim result in App.~\ref{app:moss_audit}: raters select a rubric level independently before the MOSS label is revealed, and they match it exactly on $0.257$ of mapped judgements (95\% CI $[0.191, 0.325]$) against a permutation baseline of $0.169$. On the same items two raters match each other on $0.275$ (95\% CI $[0.220, 0.337]$), so the paired human--MOSS versus human--human difference is $+0.018$ (95\% CI $[-0.053, +0.102]$) and contains zero. A rater therefore agrees with the MOSS level about as closely as two raters agree with one another, which bounds the annotation bias rather than establishing that the labels are correct; the audit covers 97 of 300 items with no complete rater pass, and the completed result will be reported whatever it shows. \voicenetext is rated by a heterogeneous eight-person pool (four with psychology degrees, contributing 95.3\% of ratings). 5,828 of its 18.5k items carry at least three ratings, but its aggregate $\kappa$ is $-0.022$ with a 95\% interval $[-0.037,-0.007]$ that now excludes zero from below, and none of the 57 attributes passes our reliable-core rule. Its present model scores are therefore preliminary and should not be used for confirmatory model ranking. Because agreement is at chance even at this coverage, additional ratings alone are unlikely to stabilise the \voicenetext rankings; a rubric revision with anchor-example rater training is required, and until then the subset is best read as an exploratory probe. Within the 40-emotion taxonomy itself, several categories elicit only weak model signal across all eleven evaluated systems --- \emph{Sourness} (taxonomy notes its primarily gustatory origin), \emph{Jealousy/Envy}, \emph{Doubt}, \emph{Interest}, and \emph{Sexual Lust} stay at $|\rho|\le0.29$ for the best column. We retain them in the released benchmark because removing them would silently shrink the taxonomy, but downstream users may wish to mask them when computing aggregate scores. The bootstrap CIs in Tab.~\ref{tab:bootstrap_ci} quantify clip-sampling uncertainty conditional on the aggregate labels and the per-prompt thresholds fitted for Tab.~\ref{tab:model_results}; they neither correct the same-set threshold optimism nor propagate rater uncertainty. Although \voiceclaplarge's paired lead is robust on all four columns, most smaller adjacent-model differences overlap zero. Four same-runtime \voiceclapsmall runs (the original seed and three exact-recipe replications) give \voicenetemo \textit{bal@pp} $0.6701\pm0.0081$ and $\rho$ $0.3122\pm0.0058$, and preliminary \voicenetext \textit{bal@pp} $0.6327\pm0.0079$ and $\rho$ $0.0986\pm0.0303$ (mean $\pm$ sample standard deviation); this is a descriptive variance proxy for Small, not evidence of seed stability for \voiceclaplarge, whose 7B-scale retraining remains infeasible within our compute budget. \voiceclapsmall and \voiceclaplarge are trained on a mixture of nine voice corpora: \emoliabal, LAION's Got Talent, Majestrino, two in-house Synthetic Vocal Bursts collections, and four FCaps-captioned corpora (EARS, Expresso, Voxceleb1, and Voxceleb2). Multilingual In The Wild is annotated and released as part of the \emolia suite, but did not improve downstream scores in an internal ablation and is therefore excluded from the released training mixture. Finally, two in-house Synthetic Vocal Bursts collections (341k clips) were curated from Creative Commons-licensed YouTube uploads; we will honour all takedown requests against the released bundle.

\section{Annotation Platform Instructions and UI}
\label{sec:annotation_platform_screenshots}
\begin{figure}[t]
    \centering
    \includegraphics[width=0.9\textwidth]{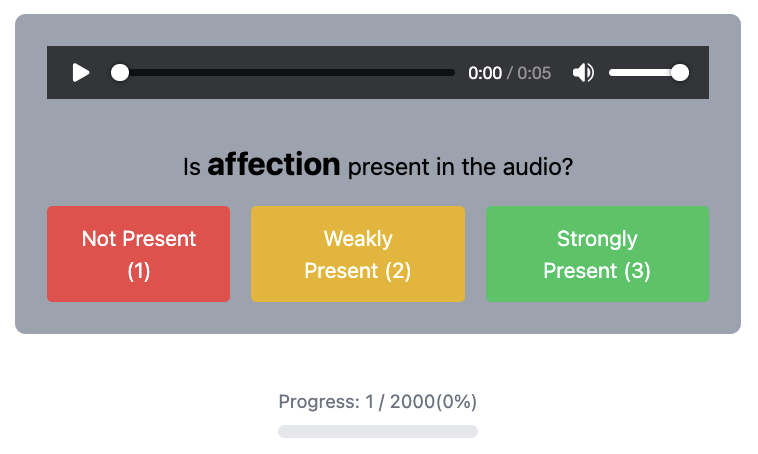}
    \caption{UI of our expert annotation tool for \voicenetemo.}
    \label{fig:annotation_platform_benchmark_annotation}
\end{figure}

\begin{figure}[t]
    \centering
    \includegraphics[width=0.9\textwidth]{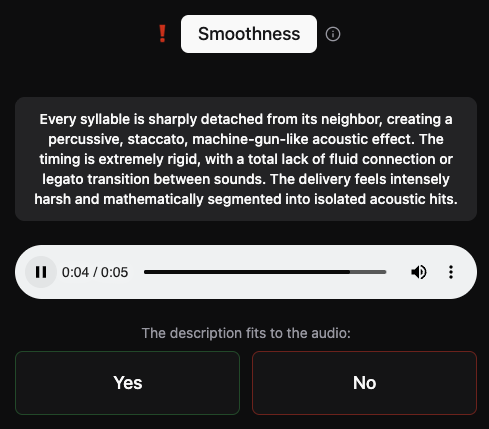}
    \caption{UI of the second annotation interface, used for the emotion-presence binary confirmation task on \voicenetemo. Annotators listen to a clip and confirm or reject whether the target emotion is present.}
    \label{fig:annotation_platform_emo_annotation}
\end{figure}

\begin{figure}[t]
    \centering
    \includegraphics[width=0.9\textwidth]{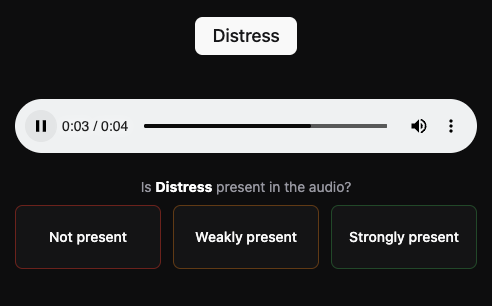}
    \caption{UI of the second annotation interface, used for the talking-style attribute binary confirmation task on \voicenetext. For each (dimension, level) bucket, annotators confirm or reject whether the audio matches the target rubric-level description.}
    \label{fig:annotation_platform_dim_annotation}
\end{figure}

Annotation platform screenshots for both benchmark subsets are shown below. The first annotation interface (Figs.~\ref{fig:annotation_platform_benchmark_instructions}--\ref{fig:annotation_platform_benchmark_annotation}) was used for the expert annotation of \voicenetemo (40-emotion rating). The second annotation interface (Figs.~\ref{fig:annotation_platform_emo_annotation}--\ref{fig:annotation_platform_dim_annotation}) was used for the binary level confirmation task on \voicenetemo and \voicenetext, respectively.

\label{annotation_platform_benchmark}
\begin{figure}[t]
    \centering
    \includegraphics[width=0.9\textwidth]{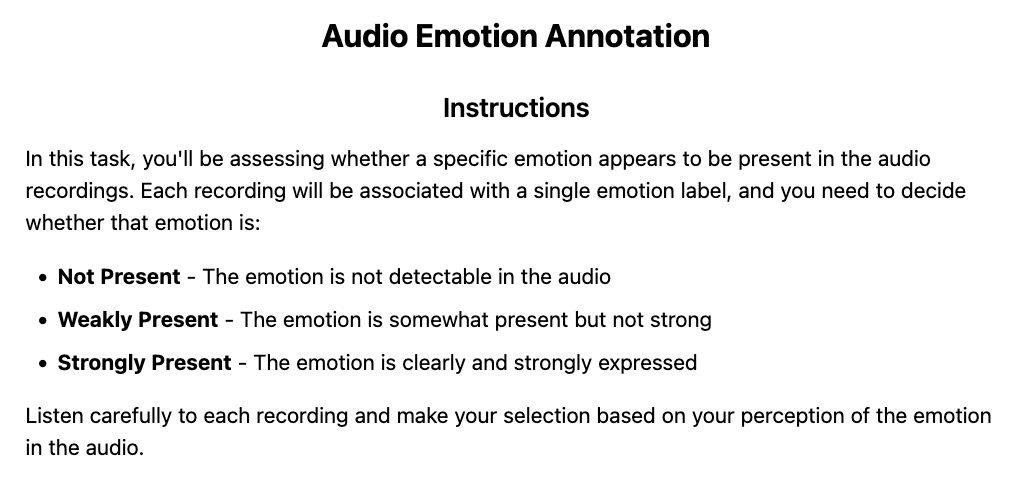}
    \caption{Instructions given to the human annotator for the expert annotation of \voicenetemo.}
    \label{fig:annotation_platform_benchmark_instructions}
\end{figure}

\end{document}